\documentclass[10pt]{iopart}

\usepackage{graphicx}
\usepackage{iopams}
\usepackage{textcomp}
\usepackage{pdfpages}

\begin{document}
	
	\title[Femtosecond spectroscopic ellipsometry]{Ultrafast dynamics of hot charge carriers in an oxide semiconductor probed by femtosecond spectroscopic ellipsometry}	
	
	\author{Steffen Richter$^{1,2,*,\mathrm{a},\mathrm{b}}$, Oliver Herrfurth$^{2,\mathrm{a},\mathrm{c}}$, Shirly Espinoza$^{1}$, Mateusz Rebarz$^{1}$, Miroslav Kloz$^{1}$, Joshua A. Leveillee$^{3}$, Andr\'{e} Schleife$^{3,\mathrm{d}}$, Stefan Zollner$^{4,5}$, Marius Grundmann$^{2}$, Jakob Andreasson$^{1}$, R\"{u}diger Schmidt-Grund$^{2,6}$}
	
	\address{$^1$ELI Beamlines/Fyzik\'{a}ln\'{i} \'{u}stav AV \v{C}R, v.v.i., Za Radnic\'{i} 835, 25241 Doln\'{i} B\v{r}e\v{z}any, Czech Republic}
	\address{$^2$Universit\"at Leipzig, Felix-Bloch-Institut f\"ur Festk\"orperphysik, Linn\'estr. 5, 04103 Leipzig, Germany}
	\address{$^3$University of Illinois, Dep. of Materials Science and Engineering, 1304 W. Green St., Urbana, IL 61801, USA}
	\address{$^4$New Mexico State University, Department of Physics, PO Box 30001, Las Cruces, NM, 88003-8001, USA}
	\address{$^5$Fyzik\'{a}ln\'{i} \'{u}stav AV \v{C}R, v.v.i., Sekce optiky, Na Slovance 2, 18221 Praha, Czech Republic} 
	\address{$^6$Technische Universit\"at Ilmenau, Institut f\"ur Physik, Weimarer Str. 32, 98693 Ilmenau, Germany}
	\address{$^\mathrm{*}$Present address: Link\"opings universitet, Institutionen f\"or fysik, kemi och biologi, 58183 Link\"oping, Sweden}
	\address{$^\mathrm{a}$These authors contributed equally.}
	\ead{$^\mathrm{b}$steffen.richter@liu.se}
	\ead{$^\mathrm{c}$oliver.herrfurth@physik.uni-leipzig.de}
	\ead{$^\mathrm{d}$schleife@illinois.edu}
	
	\newpage

	\vspace*{2.5cm}
	
	\hfill\includegraphics[height=5.5cm]{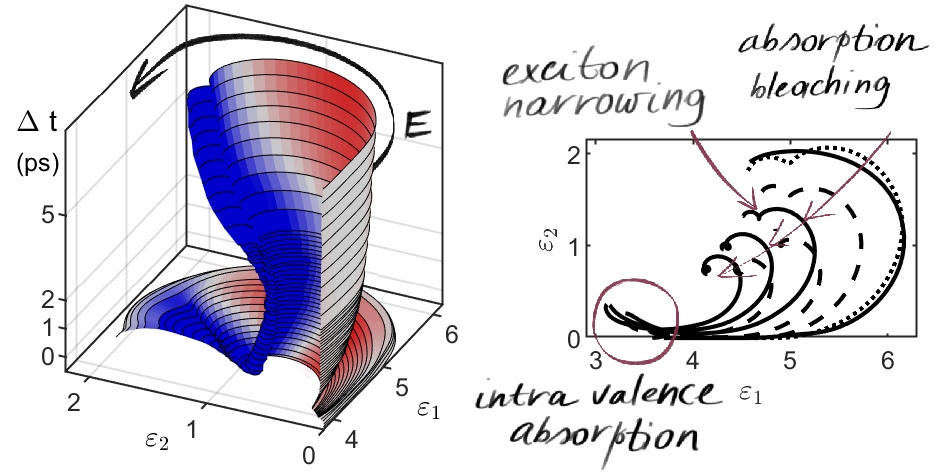}

	\vspace{.5cm}

	\begin{abstract}
		Many linked processes occur concurrently in strongly excited semiconductors, such as interband and intraband absorption, scattering of electrons and holes by the heated lattice, Pauli blocking, bandgap renormalization and the formation of Mahan excitons. In this work, we disentangle their dynamics and contributions to the optical response of a ZnO thin film. 
		Using broadband pump-probe ellipsometry, we can directly and unambiguously obtain the real and imaginary part of the transient dielectric function which we compare with first-principles simulations. 
		We find interband and excitonic absorption partially blocked and screened by the photo-excited electron occupation of the conduction band and hole occupation of the valence band (absorption bleaching). 
		Exciton absorption turns spectrally narrower upon pumping and sustains the Mott transition, indicating Mahan excitons. 
		Simultaneously, intra-valence-band transitions occur at sub-picosecond time scales after holes scatter to the edge of the Brillouin zone. 
		Our results pave new ways for the understanding of non-equilibrium charge-carrier dynamics in materials by reliably distinguishing between changes in absorption coefficient and refractive index, thereby separating competing processes. 
		This information will help to overcome the limitations of materials for high-power optical devices that owe their properties from dynamics in the ultrafast regime.
	\end{abstract}

	\vfill

	\noindent\textit{\footnotesize The abstract figure shows the transient dielectric function as Cole-Cole diagrams. Color in the 3D plot illustrates the change in $\varepsilon_2$, using a symmetric color scale similar to Fig.\,\ref{fig:psi-delta} (limits chosen as $\pm0.3$). The individual Cole-Cole diagrams on the right depict various pump-probe delay times. The line style follows Fig.\,\ref{fig:W3806_DFselectedTimes}, i.e. solid lines represent the arising pump-induced effect, dashed lines its relaxation, and the dotted line refers to the situation after 2\,ns.}


	
	\maketitle

\section{Introduction}

\noindent Many-body systems under non-equilibrium conditions, for instance caused by photo-excitation, still challenge the limits of our understanding at microscopic length and ultrashort time scales \cite{Chemla2001,Huber2001,Fleming2008}.
Accessing and controlling emergent states experimentally constitutes one of the most exciting, but also challenging, forefronts of contemporary materials science \cite{Baldini2017,Winkler2017}. 
In addition to advancing the fundamental understanding of exotic quantum states, e.g., involving large densities of free charge carriers \cite{Nenstiel2016,Versteegh2012collexon}, 
understanding such many-body systems supports technological breakthroughs and the development of novel applications including 
high-speed optical switching \cite{Colman2016,Chai2017} 
and computing \cite{Mashiko2016,Athale2016}, 
fast transparent electronics \cite{Ohta2004,Frenzel2010}, 
light harvesting \cite{Ponseca2017,Kahmann2019}, 
or even new means of propulsion for spacecrafts \cite{Atwater2018}. 
The implementation of such next-generation devices requires development of techniques that probe transient states of matter and precisely control ultrafast dynamics of excited electronic systems in solids. 

\begin{figure}[b]
	\centering
	\includegraphics[width=\textwidth]{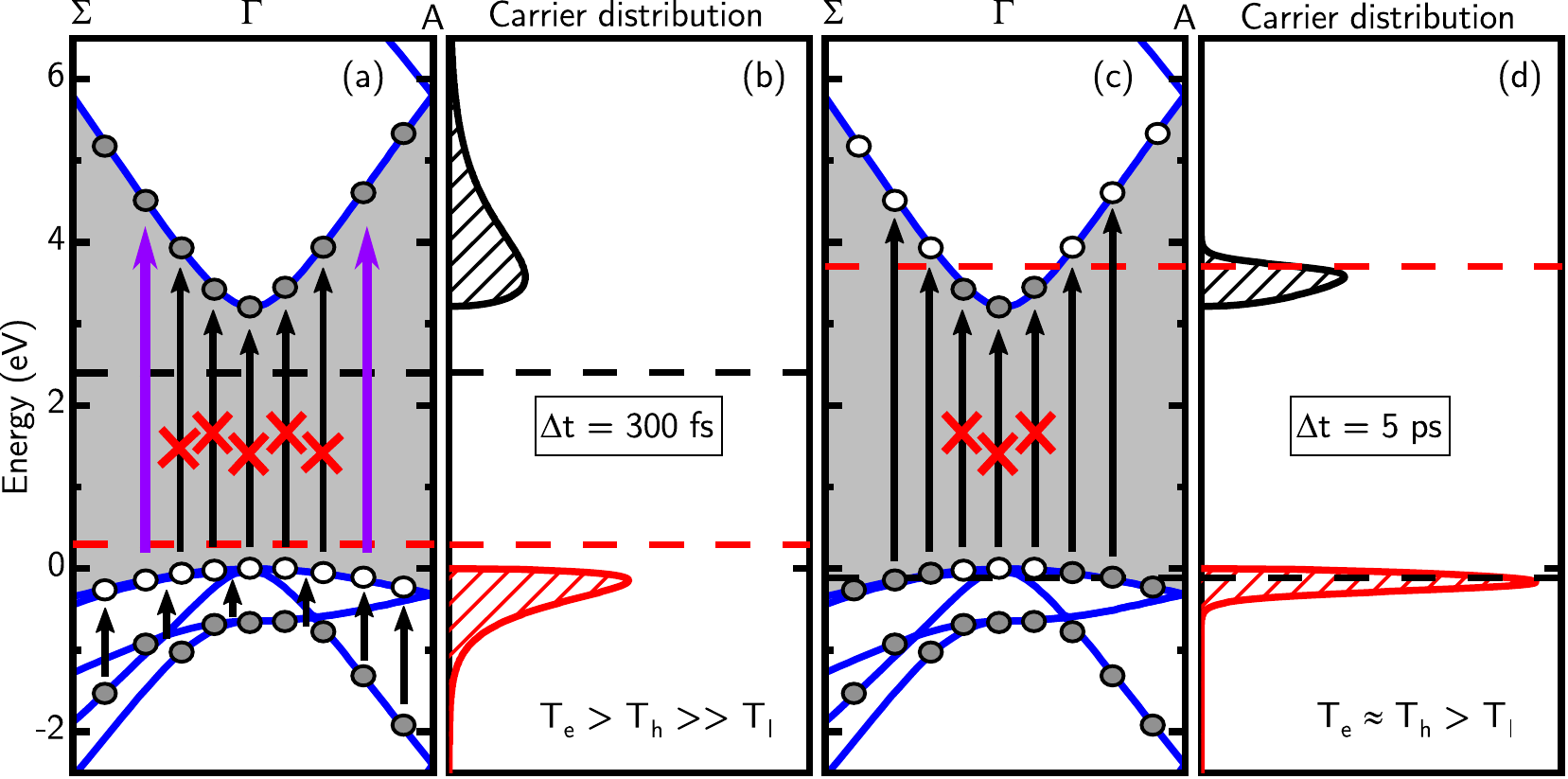}
	\caption{Hot charge carriers after strong excitation of ZnO with a UV pump pulse: 
		(a,\,b): Within a few 100\,fs after excitation (violet arrows), scattering between charge carriers results in the conduction band being occupied by excited electrons (filled circles), and the valence band by holes (open circles), (a). 
		The thermal distribution (Fermi-Dirac statistics) of the excited electrons (black) and holes (red) corresponds to effective temperatures $T_e$, $T_h$ of a few 1000\,K, (b). The quasi Fermi-energies (dashed lines) are shifted into the bandgap due to the high temperatures. 
		(c,\,d): Within the first picoseconds, scattering between charge carriers and phonons as well as recombination yield cooling and reduction of the density of excited electrons and holes. Still, charge-carrier temperatures are larger than the lattice temperature $T_{\mathrm{l}}$.   
		Black arrows in (a) and (c) mark selected optical transitions which are dynamically blocked (band-band transitions) or enabled (intra-valence-band transitions). 
	}
	\label{fig:carrierdistribution0}
\end{figure}

Many experimental and theoretical studies have aimed to separate fundamental electron-electron and electron-phonon effects, as well as the role of defect states in solids \cite{mazur-review}. 
However, due to their complexity, our understanding of the coupling between fundamental electronic excitations and the lattice remains vague, especially directly after strong excitation, i.e., on short times scales after 
electron densities as high as $10^{20}$cm$^{-3}$ have been pumped into the conduction band by a fs laser pulse. 
As illustrated schematically in Fig.\,\ref{fig:carrierdistribution0}, not only the density of excited charge-carriers (as mostly probed in luminescence experiments) matters in such a highly excited regime, but also the excess energy and the distribution of charge carriers, as well as the transient electronic band structure. Therefore, the optically accessible states change rapidly, 
enabling or prohibiting certain absorption channels, but also changing other material properties like the index of refraction. 
This is especially significant for applications of transparent semiconductors like wide-gap oxides \cite{Lorenz2016}. 

Experimentally, angular-resolved photo-electron spec\-tros\-co\-py is one of the most insightful probes for the dispersion of populated electronic states \cite{Mor2017,Zong2018}. 
Beyond this, optical spectroscopy accesses a convolution of joint density of states, electron and hole populations, and transition matrix elements via the complex, frequency-dependent dielectric function (DF). 
Conventional transient spectroscopy has been performed at different spectral ranges \cite{Doenges2016,Zuerch2017,Ziaja2015,Eisele2014}. 
Also transient sum-frequency generation was demonstrated to probe the dynamics of electronic transitions after excitation \cite{Foglia2016}. 
A general challenge is to achieve not only high time-resolution but to discriminate different processes triggered by the excitation \cite{mazur-review,Shah1999}. 
In order to provide this, one must understand the entire, i.e., spectral and complex-valued, response of an excited material. This requires obtaining both, amplitude and phase information, of a sample's DF, as encoded in its \textit{complex} reflection coefficient $\mathbf{r}$. 

Conventional transient spectroscopy yields only amplitudes and experimental data is often explained by changes in the extinction coefficient $\kappa$, neglecting changes of the refractive index $n$. This approach represents a challenge for excitation spectroscopy that has been discussed already in the 1980's \cite{Bohnert1980}. 
One way to compensate for the lack of phase information is to introduce restrictive model assumptions. 
Alternative methods are to combine measurements from different angles of incidence \cite{Huang1998,Roeser2003,Shih2009} or $p$- and $s$-polarization \cite{Boschini2015,Baldini2016}. 
However, also these methods are only work-arounds and cannot directly yield phase information. 
Alternative approaches to obtain the full dielectric response are heterodyne detection schemes \cite{Hiramatsu2015} or time-domain spectroscopy. The latter, however, is only possible in the THz regime and not at optical frequencies \cite{Poellmann2015}.

In ellipsometry, the angles $\Psi$ and $\Delta$ offer relative, frequency-dependent amplitude and phase information for the physical response, $\mathbf{r_\mathrm{p}}/\mathbf{r_\mathrm{s}}=\tan(\Psi)e^{i\Delta}$ (where indices refer to $p$- and $s$-polarizations). 
This provides simultaneous access to the real and imaginary part of the DF $\varepsilon=\varepsilon_1+i\varepsilon_2=(n+i\kappa)^2$. 
In this article, we use pump-probe spectroscopic ellipsometry to obtain transient DF spectra of photo-excited ZnO with femtosecond time-resolution. 
With its wide bandgap and excitons stable at room temperature \cite{ZnO-Buch}, ZnO is an ideal testbed for this research. 
In particular, due to its strong polarity, the strong electron-phonon coupling also impacts exciton dynamics \cite{Shokhovets2008,Oki2019}.

Our ellipsometric approach based on polarization-resolved reflectance-difference measurements gives unambiguous access to the time-dependent DF of the ZnO film after excitation with about 100\,fs temporal bandwidth. 
The experimental results yield information on the ultrafast dynamics of electron-electron and electron-phonon processes in this prototype oxide-semiconductor. 
They are complemented by first-principles simulations. 
This allows us to separate pump-induced Pauli blocking of absorption from bandgap renormalization (BGR). 
From our analysis we also report the direct observation of intra-valence-band (IVB) absorption. 
Finally, non-vanishing excitonic absorption enhancement questions the Mott transition and hints at the existence of Mahan excitons in photo-excited semiconductors. 

Our experiments improve on earlier ellipsometric pump-probe studies which suffered from shortcomings such as changing positions of the probe spot on the sample or stability issues with the femtosecond lasers, or were performed only at single wavelengths (including imaging mode) \cite{Choo1993,Zollner1997,Yoneda2003,Kruglyak2005,Mounier2008,Min2010,Rapp2016,csontos2017,Pflug2018}. 

\section{Methods}

We used a $c$-plane oriented ZnO thin film grown by pulsed laser deposition on a fused silica substrate. 
The film thickness of 30\,nm is sufficient to maintain bulk properties. 
Only a very small excitonic enhancement due to the confinement in the thin layer is expected \cite{Pal2017,Samarasingha2020}. 
At the same time 30\,nm is thin enough to assume homogeneous excitation by a 266\,nm pump pulse. 
We therefore do not need to consider the ambipolar diffusion of hot charge carriers. 
We estimate the excited electron-hole pair density to approx. $1\times10^{20}$\,cm$^{-3}$. 
The experiment is performed at room temperature.

\subsection{Time-resolved spectroscopic ellipsometry} 

\begin{figure}[b]
	\centering
	\includegraphics[width=0.80\textwidth]{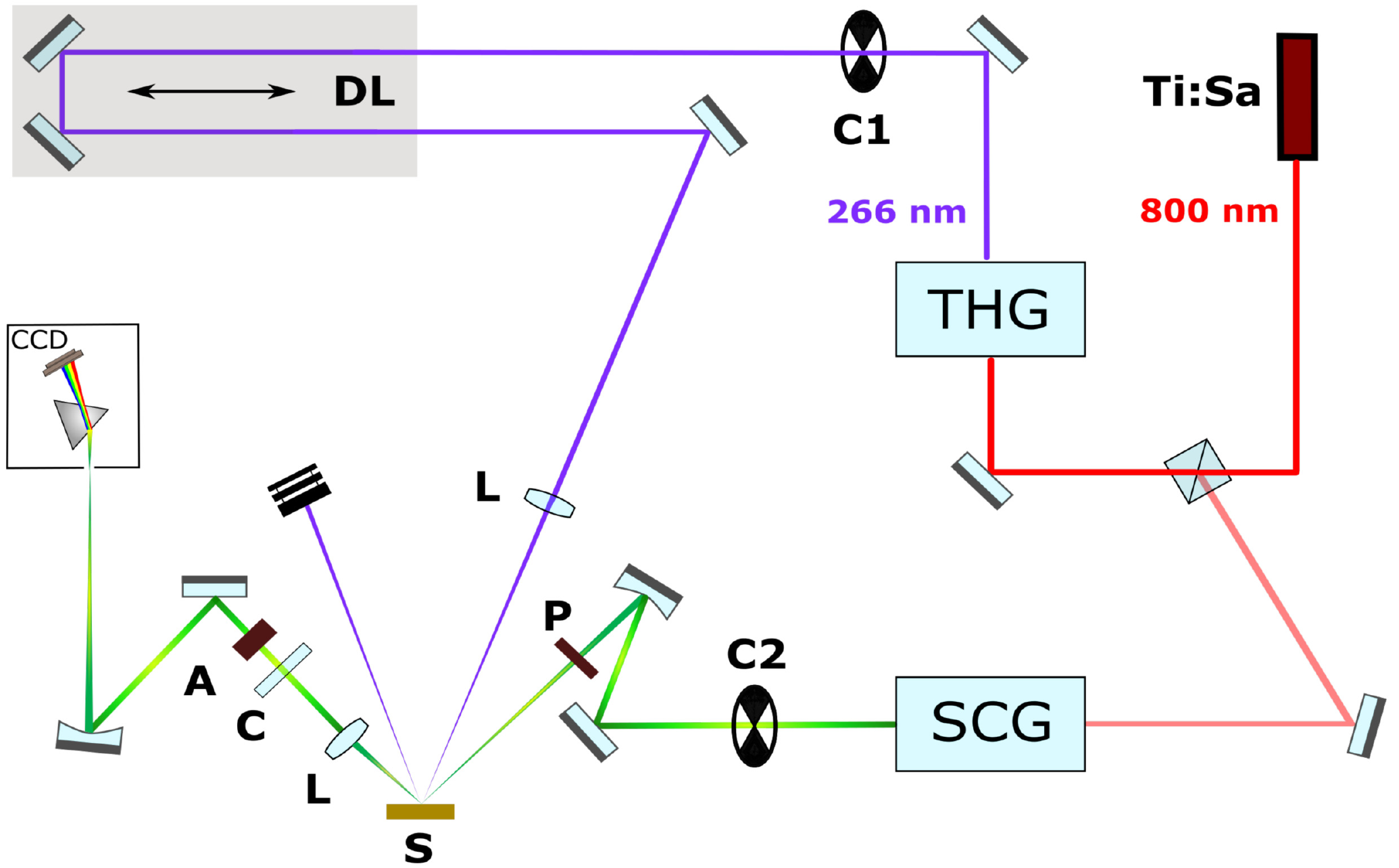}
	\caption{Schematic of the femtosecond time-resolved spectroscopic ellipsometry setup. The pump beam (\textbf{THG}) is guided through a chopper wheel \textbf{C1} ($f_{1}=250\,\mathrm{Hz}$) to a delay line \textbf{DL} and focussed by a lens \textbf{L} onto the sample. The white-light probe (\textbf{SCG}) passes through another chopper wheel \textbf{C2} ($f_{2}=500\,\mathrm{Hz}$) and is focussed onto the sample \textbf{S} by a spherical mirror through the polarizer \textbf{P}. The reflected light is collimated by a lens \textbf{L} and guided through a compensator \textbf{C} and an analyzer \textbf{A} to a prism spectrometer.}
	\label{fig:setup}
\end{figure}

We employ pump-probe spectroscopic ellipsometry, using a femtosecond pulsed laser. 
A schematic of the setup is shown in Fig.\,\ref{fig:setup}. Further descriptions can be found in Refs.\,\cite{utilitymodel,Espinoza2019}. 
The fundamental mode of a titanium sapphire laser (\textbf{Ti:Sa}) (Coherent Astrella: 35\,fs, 800\,nm, 1\,kHz repetition rate) is used for third harmonic generation (\textbf{THG}, 266\,nm), employed as pump beam. 
1\% of the laser power is used to generate supercontinuum white-light (\textbf{SCG}) in a CaF$_{2}$ window. 
In \textit{Polarizer-Sample-Compensator-Analyzer} configuration, we measure the transient reflectance-difference signal $(\Delta R/R)_j$ at 60$^\circ$ angle of incidence by scanning the pump-probe delay. 
We repeat this for ten different azimuth angles $\alpha_j$ of the compensator while the polarizer and analyzer are kept fixed at $\pm45^\circ$. \\
Spectra were captured using a prism spectrometer and a kHz-readout CCD camera (Ing.-B\"uro Stresing). 
Most critical is the fluctuating probe spectrum and amplitude due to the CaF$_2$ crystal movement as well as warm-up effects at the CCD camera. Both of these problems affect the measurement mostly on time scales larger than a few milliseconds. 
A two-chopper scheme in the pump and probe paths is employed which allows us to obtain a wavelength-dependent live-correction for the pump-probe as well as only-probe intensity spectra, hence obtaining proper reflectance-difference spectra. 
In order to compute the ellipsometric parameters from the series of measurements at different compensator angles, we apply the M\"uller matrix formalism for each photon energy and delay time, where the obtained reflectance-difference spectra are applied to reference spectra. See supplementary information for further details \cite{suppl}. 
The technique is comparable with a multi-channel lock-in system and enables comparison of spectra even if they have been measured long time after each other. Furthermore, it minimizes systematic errors from polarization uncertainties.

In order to minimize chirping of the probe pulse, we use a thin, broadband wire-grid-polarizer (Thorlabs) before the sample and focus the probe beam by a spherical mirror.  
Reflected light is analyzed by an achromatic quarter-wave plate and Glan-type prism (both B. Halle Nachfolger). 
The remaining chirp (few 100\,fs difference between 2.0\,eV and 3.6\,eV - corresponding to roughly 3\,mm dispersive material) induced by the CaF$_2$ as well as the support of the wire grid polarizer \cite{Boschini2015} is removed retroactively by shifting the zero-delay in the data analysis using an even polynomial for its wavelength dependence. 
Further details can be found in the supplementary material \cite{suppl}. \\
The focused probe spot at the sample had a $1/e^2$ diameter of 200\,\textmu m, the pump spot 400\,\textmu m (40$^\circ$, $s$-polarized) such that lateral carrier diffusion becomes negligible \cite{Herrfurth2019}. 
The corresponding temporal and spectral bandwidths are estimated to 100\,fs and 5\,nm in the UV, respectively. 

Modeling of the ellipsometry data to obtain the material's DF is performed using a transfer-matrix formalism \cite{Schubert1996} with the DF of ZnO parametrized by a Kramers-Kronig consistent B-spline function \cite{Johs2008}. 
In the model, the film is assumed to be isotropic because the experimental configuration is mostly sensitive to the DF for ordinary polarization \cite{Shokhovets2010}. 
The model is fitted to the Mueller matrix elements $N,\,C,\,S$ accounting also for spectral bandwidth and depolarization. 
The number of spline nodes was minimized in order to capture all spectral features but avoid overfitting and artificial oscillations \cite{Likhachev2017}. 
See supplementary material for error estimation based on Monte-Carlo simulations \cite{suppl}. 	 

\subsection{First-principles simulations of excited electron-hole pairs at finite temperature} 

We use first-principles simulations based on many-body perturbation theory to study the influence of electron-hole excitations on the optical properties of ZnO theoretically. 
To this end, we compute Kohn-Sham states and energies within density functional theory (DFT) \cite{Hohenberg_1964,Kohn_1965} and use these to solve the Bethe-Salpeter equation (BSE) for the optical polarization function \cite{Onida_2002}. 
All DFT calculations are carried out using the Vienna \emph{Ab-Initio} Simulation Package \cite{Gajdos:2006,Kresse:1999,Kresse:1996} (VASP) and the computational parameters described in Refs.\,\cite{Schleife2009,Schleife2011}. 
The influence of quasi-particle (QP) corrections on the band gap is taken into account using a scissor operator as also described in Refs.\,\cite{Schleife2009,Schleife2011}. 
All BSE calculations are performed using the implementation described in Refs.\,\cite{Roedl:2008,Fuchs:2008_b}. 

We compute the dielectric function as the sum of valence-conduction-band transitions $\varepsilon_{\mathrm{DFT+QP}(N,T)}^\mathrm{VBCB}$, transitions from lower valence bands into excited hole states near the valence-band maximum $\varepsilon_{\mathrm{DFT}(N,T)}^\mathrm{IVB}$, and transitions from excited electron states near the conduction-band minimum into higher conduction bands $\varepsilon_{\mathrm{DFT}(N,T)}^\mathrm{ICB}$, all depending on temperature $T$ and density of excited electrons/holes $N$:
\begin{eqnarray}
\label{eq:df_t}
\varepsilon\,(N,T,E) \\\approx  \,\varepsilon_{\mathrm{DFT+QP}(N,T)}^\mathrm{VBCB}(E)+\varepsilon_{\mathrm{DFT}(N,T)}^\mathrm{IVB}(E)
+\varepsilon_{\mathrm{DFT}(N,T)}^\mathrm{ICB}(E)+\Delta\varepsilon_\mathrm{exc}(N,E) \nonumber
\end{eqnarray}
To compute the first three contributions, we use the in\-de\-pen\-dent-particle approximation, based on ground-state electronic structure and optical dipole matrix elements, and account for temperature and excitation-density dependence by means of Fermi occupation numbers of electrons and holes. 
Thus, Burstein-Moss shift (BMS) due to Pauli blocking is automatically included in this description. 
In addition, we use the model given by Berggren and Sernelius \cite{Berggren:1981,Wu:2002} for doped systems at zero temperature to include the effect of BGR. 
BGR arises as a many-body effect due to free charge-carriers in the optically excited state and leads to a reduction of the bandgap that we include when computing $\varepsilon_{\mathrm{DFT+QP}(N,T)}^\mathrm{VBCB}(E)$. 
For a charge-carrier density of $10^{20}$\,cm$^{-1}$ by n-type doping, about 300\,meV shrinkage is assumed \cite{Kronenberger:2012}. 
When accounting for excitonic effects in the presence of high-temperature carriers, Fermi-distributed occupation numbers of electrons and holes need to be included also in the solution of the Bethe-Salpeter equation of many-body perturbation theory. 
However, this turns the eigenvalue problem for the excitonic Hamiltonian into a generalized eigenvalue problem \cite{Bechstedt:2015}. 
Here, we avoid this complication and increase in computational cost and, instead, approximate excitonic effects using the zero temperature difference $\Delta\varepsilon_\mathrm{exc}(N,E)$,
\begin{equation}
\label{eq:bse_n}
\Delta\varepsilon_\mathrm{exc}(N,E) = \varepsilon_{\mathrm{BSE+QP}(N)}(E)-\varepsilon_{\mathrm{DFT+QP}(N)}(E).
\end{equation}
Here, $\varepsilon_{\mathrm{BSE+QP}(N)}(E)$ is the dielectric function with excitonic effects and $\varepsilon_{\mathrm{DFT+QP}(N)}(E)$ is the corresponding independent-quasi-particle dielectric function, the band gaps for both are corrected using a scissor operator \cite{Schleife2009,Schleife2011}.
While this approach neglects the influence of temperature on excitonic effects, both $\varepsilon_{\mathrm{BSE+QP}(N)}(E)$ and $\varepsilon_{\mathrm{DFT+QP}(N)}(E)$ include BGR and BMS.

To compute $\varepsilon_{\mathrm{BSE+QP}(N)}(E)$, we extend the framework described in detail in Refs.\,\cite{Schleife2011,Schleife:2011_b,Kang:2019} to describe excited electrons (e) and holes (h) at a temperature of zero K. 
Here, the lowest conduction band states are occupied with free electrons of density $N$, and the highest valence states with holes of the same density $N$. 
Hence, transitions between these states are excluded. This is described in our framework via occupation numbers of otherwise unchanged single-particle Kohn-Sham states when computing the BSE Hamiltonian. 
Finally, our framework accounts for electronic interband screening of the electron-hole interaction in the BSE Hamiltonian, using the static dielectric constant obtained in in\-de\-pen\-dent-particle approximation, $\varepsilon_\mathrm{eff}$=4.4\,. 
In addition, as discussed earlier for doped ZnO \cite{Schleife2011}, excited carriers modify the electron-hole interaction by contributing intraband screening. 
We approximate this contribution using the small-wave-vector limit of a static, wave-vector ($q$) dependent Lindhard dielectric function, which, in the presence of free electrons and holes becomes\cite{Schleife2011,Schleife:2011_b,Kang:2019}
\begin{equation}
\label{eq:epsqtf}
\varepsilon_\mathrm{intra}(q) \approx 1 + \frac{q_\mathrm{TF,e}^2}{q^2}+\frac{q_\mathrm{TF,h}^2}{q^2},
\end{equation}
with the Thomas-Fermi (TF) wave-vectors 
\begin{equation}
\label{eq:qtf}
q_\mathrm{TF,e/h}=\sqrt{\frac{3Ne^2}{2\varepsilon_0\varepsilon_\mathrm{eff}\tilde{E}_\mathrm{F}^\mathrm{e/h}}}.
\end{equation}
The relative Fermi energies of electrons and holes at zero temperature, $\tilde{E}_\mathrm{F}^\mathrm{e/h}$,
\begin{equation}
\label{eq:fermi}
\tilde{E}_\mathrm{F}^\mathrm{e/h}=\frac{\hbar^2}{2m_\mathrm{e/h}}\left(3\pi^2N\right)^{2/3},
\end{equation}
refer to the conduction-band minimum and valence-band maximum,  $\tilde{E}_\mathrm{F}^\mathrm{e}=E_\mathrm{F}^\mathrm{e}-E_\mathrm{CB}$ and $\tilde{E}_\mathrm{F}^\mathrm{h}=E_\mathrm{VB}-E_\mathrm{F}^\mathrm{h}$, respectively. 
Equation \ref{eq:epsqtf} then becomes
\begin{equation}
\label{eq:epsqtf2}
\varepsilon_\mathrm{intra}(q) = 1 + \frac{1}{q^2} \frac{3N e^2}{2\varepsilon_0\varepsilon_\mathrm{eff}}\left( \frac{2\left(m_\mathrm{e}+m_\mathrm{h}\right)}{\hbar^2}\frac{1}{\left(3\pi^2N\right)^{2/3}}\right).
\end{equation}
$\varepsilon_\mathrm{intra}$ enters the $W$ term in the exciton Hamiltonian that is used to compute $\varepsilon_{\mathrm{BSE+QP}(N)}$. 
Effective electron and hole masses are parametrized using parabolic fits to our first-principles band-structure data, leading to $m_\mathrm{e}$=$0.3 m_0$.
For the hole effective mass in Eq.\,\ref{eq:epsqtf2} we use the geometric average of the masses of the three degenerate uppermost valence bands, i.e.\ $m_\mathrm{h}$=$0.62 m_0$.
This approach is valid for zero temperature of the free carriers and its implementation in our BSE code \cite{Schleife2011} allows us to compute the dielectric function, including excitonic effects, as a function of free-charge-carrier concentration $N$.

Finally, to compare with experimental pump-probe data, we compute and visualize the difference
\begin{equation}
\label{eq:df_diff}
\Delta\varepsilon = \varepsilon(N,T,E) - \varepsilon_\mathrm{BSE+QP(N=0)}(T\!=\!0\!\mathrm{~K},E).
\end{equation}

\section{Experimental results}

\begin{figure}[t]
	\centering
	\includegraphics[width=\textwidth]{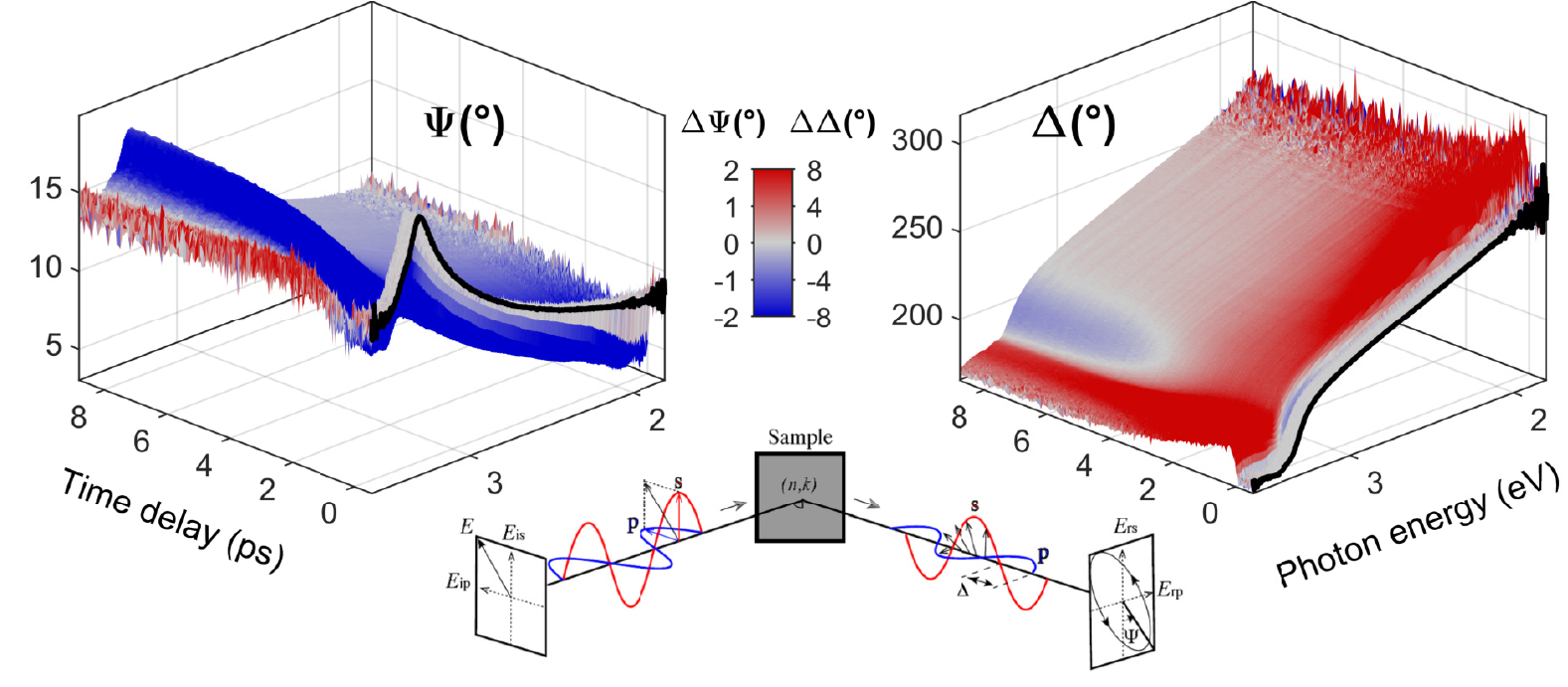}
	\caption{Temporal evolution of the ellipsometric angles $\Psi$ (amplitude ratio) and $\Delta$ (phase difference) of the ZnO thin film after non-resonant UV pump measured at 60$^\circ$ angle of incidence. 
		Increases relative to the initial spectra before excitation (black) are shown in blue, decreases in red. 
		The sketch at the bottom illustrates the meaning of the ellipsometric parameters.
	}
	\label{fig:psi-delta}
\end{figure}

The experiments were performed on a ZnO thin film, pum\-ped by 266\,nm, 35\,fs laser pulses that created an electron-hole pair density of $10^{20}$\,cm$^{-3}$. 
Supercontinuum white-light pulses were used as a probe. 
The transient ellipsometric angles $\Psi$ and $\Delta$ obtained in the spectral range 1.9-3.6\,eV are shown in Fig.\,\ref{fig:psi-delta}. 
The uncertainty of the transient changes depends on the wavelength but is mostly on the order of 0.03$^\circ$ for $\Psi$ and 0.2$^\circ$ for $\Delta$. 
See supplementary material for comprehensive error estimation \cite{suppl}. 
Figure \ref{fig:psi-delta} shows that the pump causes an immediate decrease in $\Psi$ and an increase in $\Delta$ with spectrally and temporally varying details. 
Data were recorded at delays up to 2\,ns with increasing delay steps. 

\begin{figure*}[tb]
	\centering
	\includegraphics[width=.9\textwidth]{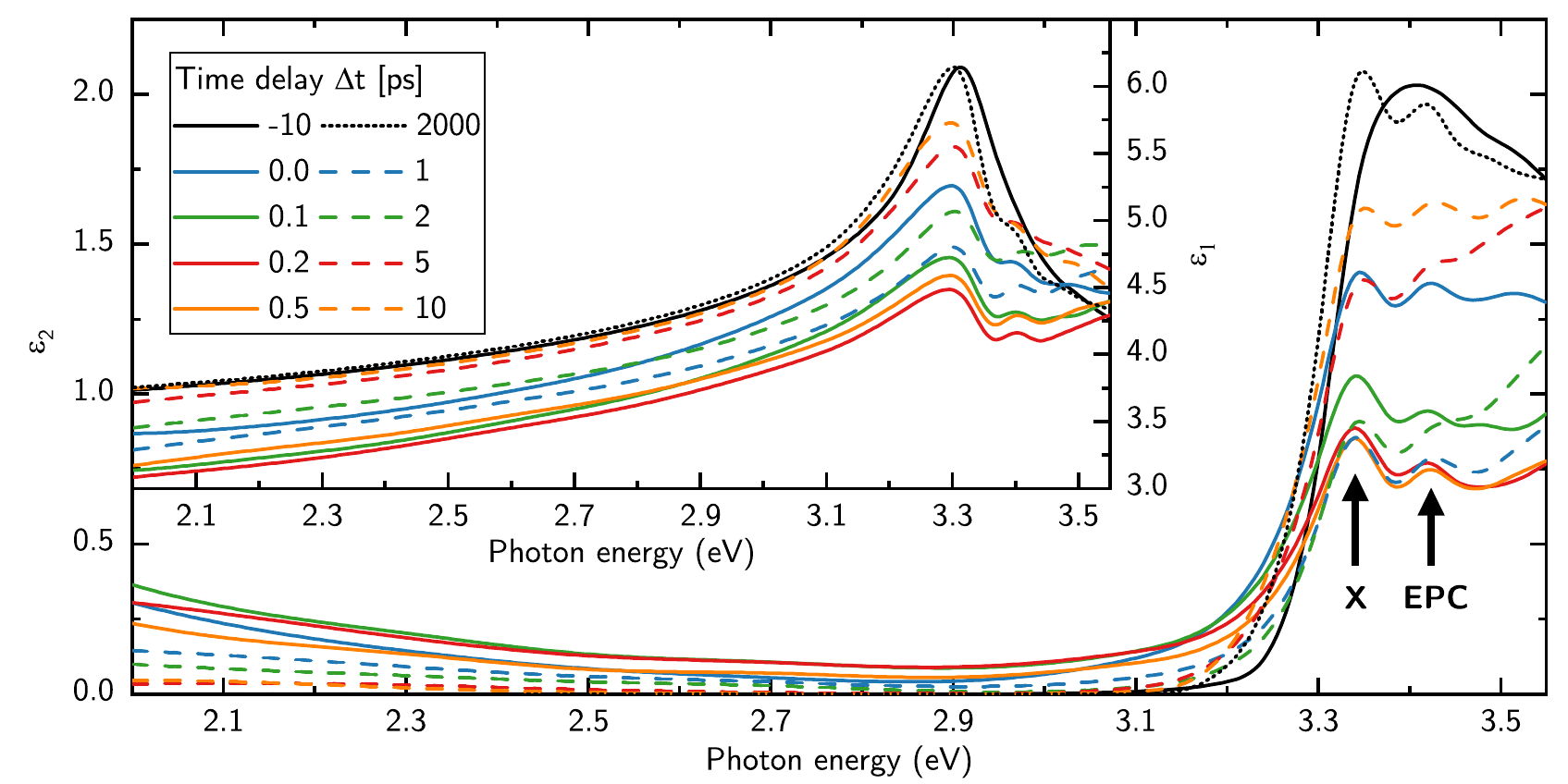}
	\caption{Real ($\varepsilon_1$, inset) and imaginary ($\varepsilon_2$, parent figure) part of the DF of the ZnO thin film at selected pump-probe delays. 	
	}
	\label{fig:W3806_DFselectedTimes}
\end{figure*}

\begin{figure}[!htb]
	\centering
	\includegraphics[width=0.7\textwidth]{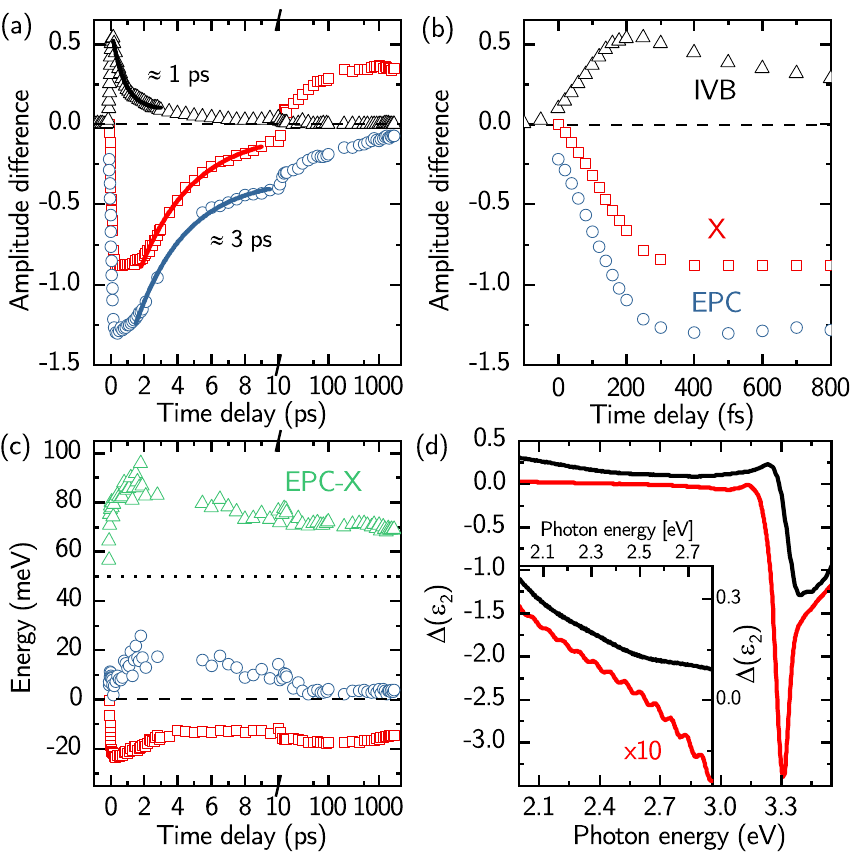}
	\caption{Evolution of absorption amplitude (a,\,b) and peak energy (c) of the exciton transition (red, X) and exciton-phonon complex (blue, EPC) as obtained from the maxima of $\varepsilon_2$ (cf.\,Fig.\,\ref{fig:W3806_DFselectedTimes}). Black symbols in (a,\,b) depict the integrated value of $\varepsilon_{2}$ in the spectral range 2.0\,eV to 3.1\,eV for different delay intervals (IVB). 
	The green symbols in (c) show the spectral difference between X and EPC which is related to an effective phonon energy $E_\mathrm{ph}$. Its equilibrium value of about 50\,meV \cite{Shokhovets2008} is indicated by the dotted line. Solid lines indicate exponential processes with their time constants. 
	(d): comparison of computed (red) and experimental (black) $\varepsilon_2$ at maximum change. In the inset, the computed IVB absorption is enhanced 10x for better display of the spectral dependence.
	}
	\label{fig:transients}
\end{figure}

From the ellipsometric spectra, we obtain the DF of the ZnO film for every pump-probe delay $\Delta t$.
Figure \ref{fig:W3806_DFselectedTimes} illustrates the resulting DF $\varepsilon=\varepsilon_{1}+i\varepsilon_{2}$ at selected delays, and Fig.\,\ref{fig:transients} represents transient evaluations. 
The uncertainty of the DF can be estimated to the order of 0.03 for $\varepsilon_1$ and 0.02 for $\varepsilon_2$, see supplementary material \cite{suppl}. 
At negative $\Delta t$, the obtained DF coincides with the one obtained in steady-state ellipsometry. 
The peak around 3.35\,eV comprises the excitonic transitions (X) and the peak around 3.42\,eV is associated with exciton-phonon complexes (EPC) \cite{Shokhovets2008}. 
There exist also further complexes at slightly higher energy. 

For small positive $\Delta t$, the absorption at the band edge and above is strongly damped, as indicated by the prominent decrease in $\varepsilon_{2}$ (Fig.\,\ref{fig:W3806_DFselectedTimes}). 
In particular, the absorption peaks of exciton and EPC are both bleached within 400\,fs (red and blue symbols in Fig.\,\ref{fig:transients}\,(b)). 
This is accompanied by a reduced refractive index below the band edge as illustrated by the concomitant drop in $\varepsilon_1$ (Fig.\,\ref{fig:W3806_DFselectedTimes}). 
Vanishing of the absorption bleaching does not start until approx. 1\,ps. 
And we note that the excitonic enhancement does not completely vanish at any time, as indicated by the dynamics of the peak structure in $\varepsilon_{2}$. 
Absorption recovery starts from higher energies, approaching the fundamental excitonic absorption peak only at later times. 
After 2\,ps both the exciton and EPC absorption peaks recover with time constants of 3\,ps, slowed down after 10-20\,ps with a non-exponential evolution (Fig.\,\ref{fig:transients}\,(a)). 
The spectral broadening of the exciton and EPC transitions was reduced as soon as the sample has been excited (Fig.\,\ref{fig:W3806_DFselectedTimes}). This reduced broadening remains approximately constant for at least 2\,ns. 

As Fig.\,\ref{fig:transients}\,(c) indicates, an immediate redshift of the exciton energy by roughly 20\,meV is followed by an increase with a linear rate of approx. 3\,meV/ps during the subsequent 4\,ps (red symbols in Fig.\,\ref{fig:transients}\,(c)).  
The EPC follows the trend with even larger increase but without the initial redshift. 
Another later redshift of both yields a minimum of exciton and EPC energies at 100\,ps. 
At 2\,ns, the absorption edge remains redshifted by approx.~20\,meV. 
It should be noted that the energetic difference between the exciton and EPC absorption peaks, which had initially increased by more than 30\,meV, approaches its steady-state value (approx.~50\,meV) monotonically until complete relaxation has happened after several nanoseconds (green symbols in Fig.\,\ref{fig:transients}\,(c)). 

Simultaneously with the arrival of the pump laser pulse, also a broad absorption arises at low energies in the bandgap (Fig.\,\ref{fig:W3806_DFselectedTimes}). 
This absorption reaches its maximum amplitude at $\Delta t=0.2$\,ps, and then decreases with a time constant of 1\,ps. It vanishes completely after 10\,ps (black symbols in  Fig.\,\ref{fig:transients}\,(a,\,b)).

\newpage

\section{Discussion}

\subsection{Separating Physical Processes}

Charge carrier excitation by 266\,nm (4.67\,eV) laser pulses in ZnO involves optical transitions from the heavy-hole, light-hole and split-off valence-bands (VB) into the conduction band (CB) in the vicinity of the $\Gamma$ point, as illustrated by the violet arrows in Fig.\,\ref{fig:carrierdistribution0}\,(a).
The excited electrons carry excess energies of almost 1\,eV, the excited holes only about 0.4\,eV because of their larger effective mass (see supplementary material \cite{suppl}). 
Hence, while many transient spectroscopy experiments reported in the literature focus on near-resonant excitations, we induce very high excess energies to the charge carriers. 
The initial occupation of electron and hole states due to the pump pulse is sharply peaked and non-thermal. It takes a few hundred femtoseconds until a Fermi-Dirac distribution is established as sketched in Fig.\,\ref{fig:carrierdistribution0}\,(a) and (b). 
This excited state is referred to as hot charge-carriers. 
Estimated effective temperatures for the electrons, holes and lattice in this state are reported in Table \ref{tab:subsystems}. 
Initial thermalization is provided mainly through carrier-carrier scattering and partially through carrier-phonon scattering \cite{mazur-review,Shah1999,Othonos1998,Gattass2008,mao2004}. 
The immediate effect of this process on the optical response spectra (Fig.\,\ref{fig:W3806_DFselectedTimes}) is three-fold: 
First, 
the occupation of the states leads to (partial) Pauli blocking (band filling) and hence the observed absorption bleaching of the band-to-band and excitonic transitions. This is illustrated by crossed-out black arrows  in Fig.\,\ref{fig:carrierdistribution0}\,(a). 
An additional reduction of the excitonic absorption enhancement is expected from free-charge-carrier screening. 
The related reduction of the refractive index in the visible spectral range results from the Kramers-Kronig relations. 
Second, 
due to the flatness of the valence bands, excited holes have enough excess energy to scatter towards the edge of the Brillouin zone and thus promote IVB transitions (Fig.\,\ref{fig:carrierdistribution0}\,(a), short black arrows) which are observed as low-energy absorption, similar to observations for strongly doped $p$-type semiconductors \cite{Segura1997,Lao2011}. 
Third, the high density of photo-excited charge carriers results in BGR as seen by the redshift of the exciton energy (Fig.\,\ref{fig:transients}\,(c), red symbols). 
Additionally, the excited carriers screen a static electric field in the film that otherwise arises from Fermi-level pinning at the surface caused e.g. by donor-like oxygen vacancies \cite{Allen2010}: 
While the observed steady-state broadening of the excitons is caused by the related band bending, 
pump-induced charge carriers reduce it by leveling out this band bending. 
At large delay times, the vacancies are still passivated by trapped electrons at the surface, thus the excitonic narrowing remains for a rather long time. 
Deeply trapped holes can last for microseconds \cite{Foglia2018}.
We also note that the excitonic narrowing is strongly connected to the nature of the polar ($c$-plane oriented) thin film, which behaves similar to a broad quantum well with incorporated static electric field \cite{Stolzel2012}. We did not observe the exciton narrowing in a non-polar $m$-plane oriented ZnO film.

\begin{figure}
	\centering
	\includegraphics[width=0.5\textwidth]{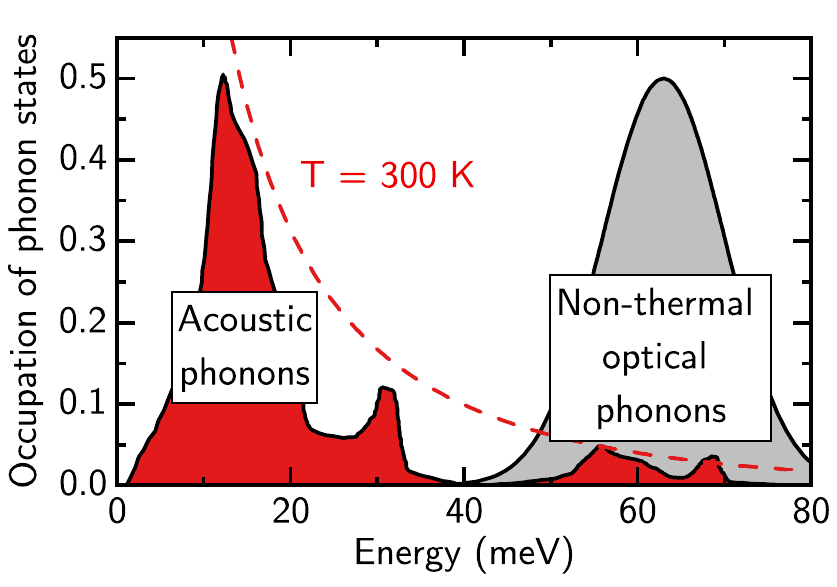}
	\caption{Simplified distribution of non-thermal (though hot) phonons after charge-carrier relaxation: 
	The strong electron-LO-phonon interaction during cooling of the charge carriers yields a non-thermal occupation of optical phonons (gray) in contrast to the occupation of, mostly acoustic, thermal phonons (red) which follows a Bose-Einstein-distribution (red dashed line) before excitation and after lattice relaxation. The phonon density-of-states is taken from \cite{Bachmann2012}.
	}
	\label{fig:phonondistribution}
\end{figure}

Analysis of the transients yields insights into individual dynamics: 
Charge-carrier thermalization is slightly faster for holes (200\,fs) than for electrons (400\,fs) because of their lower excess energy. 
This is observed in the experiment by a slightly shorter rise time until maximum response for the IVB absorption compared to the exciton bleaching (cf.\,Fig.\,\ref{fig:transients}\,(b)). 
The subsequent fast decay of the IVB absorption (Fig.\,\ref{fig:transients}\,(a)), is a consequence of the hole occupation far from the $\Gamma$ point. 
Hence, its $1/e$ decay time of 1\,ps reflects mainly the hole cooling by scattering with phonons. 
Also the transient dynamics of the absorption bleaching is governed by mainly the decrease of electron and hole temperatures: 
Initially, electrons have much higher excess energy and hence higher effective temperature than holes (Tab.\,\ref{tab:subsystems}). 
Over time, they approach each other, resulting in the situation sketched in Fig.\,\ref{fig:carrierdistribution0}\,(c,\,d).  
It is known that the carrier cooling occurs mainly by (polar) scattering with optical phonons \cite{Shah1999,Oki2019,Shah1978,Hess1999}, and effective charge-carrier temperatures do not decrease mono-exponentially \cite{Sadasivam2017}. 
The underlying electron-LO-phonon (Fr\"{o}hlich) interaction is generally very fast in polar materials, for ZnO with a Fr\"ohlich constant $\alpha$ of about 1.2 \cite{Grundmann2016} and $E_\mathrm{LO}\approx$72\,meV, the classical scattering rate is on the order of $10^{14}$\,Hz \cite{Chen2004,DasSarma1988}. Hence, cooling of the charge-carriers should take not longer than $\approx$\,0.5\,ps \cite{Shah1999}. 
However, here, the high excess energy of the charge carriers causes an extraordinarily large population of LO phonon states upon scattering, and thus an intermediately non-thermal phonon distribution as sketched in Fig.\,\ref{fig:phonondistribution}. 
A lattice temperature is not even well defined at this state. 
These excess phonons slow down the electron relaxation through phonon re-absorption by the charge carriers, resulting in the plateau-like transient during the first 2\,ps where the relaxation is delayed (Fig.\,\ref{fig:transients}\,(a), blue and red symbols). This is referred to as hot-phonon effect and was observed earlier \cite{Shah1999,Othonos1998,Poetz1983,vonderLinde1980}. 

In general, the presence of many charge-carriers can also screen the electron-phonon interaction itself \cite{DasSarma1988,Sjodin1998,Yoffa1981,Seymour1982}. 
Screening can slow down the relaxation by affecting both the Fr\"ohlich coupling constant and the LO phonon energy \cite{Ehrenreich1959,Alfano}. 
For $10^{20}$cm$^{-3}$ excited electron-hole pairs, we can estimate an effective plasma energy of $\hbar\omega_\mathrm{p}\approx0.45$\,eV (parabolic approximation, see supplementary material \cite{suppl}). 
This value exceeds the LO phonon energy by far. 
Hence we do not expect a dramatically altered (reduced) electron-phonon interaction as it would occur for $\hbar\omega_\mathrm{p}\approx E_\mathrm{LO}$ \cite{Ehrenreich1959,Leheny1979,Shah1999}. 
The fact that for electrons and holes with very large excess energy, screening is not the most dominating effect in their relaxation dynamics, is also consistent with the finding that the excitonic absorption peaks do not vanish entirely (see discussion below). 
Finally, also the plateaus of the absorption amplitudes during the first 2\,ps (Fig.\,\ref{fig:transients}\,(a), blue and red symbols) hint at a saturation related to hot phonons. 
However, the hot-phonon effect may be reduced by charge-carrier screening \cite{Shah1999}. 

The non-thermal phonon distribution is also observed by the increased energetic splitting between exciton and EPC (Fig.\,\ref{fig:transients}\,(c), green symbols): 
The effective absorption peak of the EPC at 3.42\,eV is expected to involve several phonons with an effective energy $E_\mathrm{ph}$ on the order of 30\,meV resulting in about 50\,meV splitting \cite{Shokhovets2008}. 
The absorption and re-emission of many optical phonons by the crystal apparently increases the interaction probability of (high-energy) optical phonons with excitons while (low-energy) acoustic phonons are effectively suppressed, i.e. $E_{\mathrm{ph}}$ increases. 
After more than 2\,ps the charge carriers have largely cooled down, and the number of non-thermal phonons has reached a maximum (see difference of exciton and EPC peak, Fig.\,\ref{fig:transients}\,(c)). \\
We observe that the return of the EPC absorption (vanishing Pauli blocking) has started already at times before 2\,ps, i.e. slightly earlier than for the exciton absorption (Fig.\,\ref{fig:transients}\,(a)). 
This is because the occupation of energetically higher EPC states decreases faster than the occupation of states at the CB minimum and VB maximum. \\
Lastly, we do not observe coherent phonon oscillations \cite{Ishioka2010,Lin2016,Baldini2019,Cho1990} because our pump energy is highly non-resonant with the band gap and well above our spectral probe window.

\begin{table}
	\caption{Statistics of the electron, hole, and lattice subsystems; situation immediately after charge-carrier thermalization following a pump laser pulse with 4.67\,eV that excited $10^{20}$\,cm$^{-3}$ electron-hole pairs in the ZnO thin film. The increase of $T_{\mathrm{l}}$ after complete equilibration is estimated to 50\,K at most. See supplementary material for details. 
		\\[-.5em]}
	\centering
	\begin{tabular}{c c c}
		\hline
		& temperature & quasi Fermi-energy \\
		\hline \vspace{-.3cm} \\
		electrons & $T_{\mathrm{e}}\approx7000$\,K & $E_{\mathrm{F}}^{\mathrm{e}}-E_{\mathrm{CB}}<-660$\,meV  \\
		holes &$T_{\mathrm{h}}\approx2800$\,K & $E_{\mathrm{VB}}-E_{\mathrm{F}}^{\mathrm{h}}<-260$\,meV \\
		lattice & $T_{\mathrm{l}} \approx ~300\,\mathrm{K}$ & \\ 
		\hline
	\end{tabular}
	\label{tab:subsystems}
\end{table}

Effects of the high charge-carrier densities can be distinguished from the thermal excess: 
A reduction of the total number of excited charge carriers is observed by the vanishing BGR within the first picoseconds (cf. exciton peak energy in Fig.\,\ref{fig:transients}\,(c)). 
In the subsequent picosecond regime, we expect also the recovery of the exciton and EPC absorption to result from the reduction of the excited carrier density. 
This recombination is initially maintained mainly by nonradiative Auger and defect recombination \cite{Ou2012}. 
We find an initial time constant of 3\,ps (Fig.\,\ref{fig:transients}\,(a)). 
At later times with lower remaining carrier densities, slower radiative electron-hole recombination becomes dominant. 

Thermal equilibration with the lattice can be estimated to be accomplished approx.~100\,ps after the excitation when the exciton energy reaches another minimum (see Fig.\,\ref{fig:transients}\,(c), red symbols) that indicates the highest achieved lattice temperature and thus bandgap shrinkage \cite{Rai2012}. 
Assuming a deposited energy density of 100\,J/cm$^3$ by the pump pulse, a maximum temperature increase of 30-50\,K can be expected. 
If transferred entirely to the lattice, this would correspond to a bandgap decrease of approx. 25-30\,meV at most. This matches the experimental observation. 
The following slow (approx. 2\,\textmu eV/ps) heat dissipation lasts until at least 10\,ns. 
It should be noted that the observed overshooting of the exciton amplitude at later time is related to the reduced exciton broadening as discussed above. 

\subsection{Interpretation}

To complement our experimental results, we use first-principles electronic-structure calculations to explain the different effects near the band edge: 
i) Many-body perturbation theory is used to describe excitonic effects. It includes additional screening and Pauli blocking due to the high density of electrons in the conduction and holes in the valence band at 0\,K temperature. 
ii) The high effective temperatures of excited electrons and holes are taken into account via Fermi-distributed occupation numbers in the absorption spectrum of non-interacting electron-hole pairs.  

A comparison with the experimental data in Fig.\,\ref{fig:transients}\,(d) shows that the observed reduction of the exciton absorption is only about half of what is expected from the calculations. 
An increased number of free charge carriers is known to have two opposing effects on the band-edge absorption: While the exciton is screened and should shift toward higher energies due to a reduced binding energy, the bandgap shrinks due to BGR. 
Both compensate each other in a good approximation, such that the absolute exciton energy remains constant \cite{Gay1971,klingshirn1981,Zimmermann1988,yamamoto1999}. 
However, when surpassing the so called Mott transition, excitons should cease to exist, and BGR should take over. 
That could explain the initial redshift (Fig.\,\ref{fig:transients}\,(c)) which has been observed earlier \cite{yamamoto1999,Shih2009,acharya2014}. 
Nevertheless we find that the excitonic absorption peak does not vanish entirely at any time. 
That reflects the difference between an equilibrated system and hot charge carriers: 
In the case of doping ZnO by $10^{20}$\,cm$^{-3}$ excess electrons, a Burstein-Moss blueshift of the absorption edge of approximately 370\,meV can be estimated from band-structures and band dispersions computed using density-functional-theory. 
It is clear that this does not apply to a hot electron-hole plasma where no strong blueshift is observed \cite{Bohnert1980,yamamoto1999,acharya2014}. 
Also BGR, which should generally not depend on temperature \cite{Klingshirn2012}, was found slightly less efficient for hot charge carriers \cite{Zimmermann1988}. 
Hence, due to the widely-distributed hot carriers, it is possible that the Mott transition does not occur despite the fact that the total density of excited charge carriers is well beyond the classical threshold \cite{Klingshirn2007}. 
According to \cite{Versteegh2011}, the fraction of photo-excited charge carriers bound to excitons is rather small, not exceeding 15\%. In this respect, the non-vanishing exciton absorption peaks could indicate that the occupation of the exciton ground-state would never exceed the Mott density even if $10^{20}$\,cm$^{-3}$ electron-hole pairs are excited. 
While comparisons of charge-carrier densities do not account for the effective carrier temperatures, the Mott transition can also be considered in terms of the Debye screening lengths \cite{Klingshirn2012}. 
Here, we obtain screening lengths on the order of 3$\mathring{\mathrm{A}}$, which is much smaller than the exciton Bohr radius of about 2\,nm in ZnO \cite{Versteegh2011}. 
Hence, effectively uncorrelated electron and hole states should dominate (see supplementary material for details \cite{suppl}). 
On the other hand, not only interaction lengths matter but also relevant electron (and hole) wave vectors must be similar to the Thomas-Fermi wave-vectors for effective screening \cite{Ehrenreich1959}. 
Considering the strong effect of non-parabolicity at the high excess energies of the photo-excited electrons and holes in our experiment, this can give rise to a reduced screening effect (see supplementary material for details \cite{suppl}). 
Albeit largely screened and broadened, electron-hole coupling has indeed been observed to sustain the apparent Mott transition as so called Mahan excitons \cite{Mahan1967,Haug1978,Zimmermann1988,Palmieri2020}, in particular also ZnO has been considered \cite{Schleife2011}. 
Narrow exciton-like peaks have been observed well above the Mott transition in highly-doped GaN \cite{Nenstiel2016}, Cooper-pair signatures at the Fermi sea in highly-excited ZnO \cite{Versteegh2012collexon}. 
We conclude that the remaining absorption peaks that we observe in our experiment are likely to be Mahan excitons \cite{Mahan1967,Schleife2011}. While they were originally studied in degenerately doped semiconductors, our photo-excitation generates at the same time electrons in the conduction band and holes in the valence band. 

The obvious explanation for photo-induced absorption at lower photon energies would be free carriers \cite{Hendry2007}. 
However, two Drude terms to describe free electrons and holes with the known densities and reasonable effective masses and mobilities cannot mimic the shape of the dispersion of the DF well. 
Furthermore, there are indications of a maximum of $\varepsilon_{2}$ around 1.9\,eV and 2.1\,eV (cf.\,Fig.\,\ref{fig:W3806_DFselectedTimes}) hinting at IVB transition at the M point. 
In a recent report, similar absorption features, induced by lower pump power and at much longer time scales, were attributed to defect states \cite{Foglia2018};  however, defects can hardly explain the large absorption cross sections ($\varepsilon_2$) that we observe here. 
Effects of defects in ZnO can vary strongly between different samples \cite{Schmidt2013}. 
To quantify absolute contributions of both effects, a detailed study of the dependence of this absorption on the photo-induced charge-carrier density would be needed but is beyond the scope of this article. However, we saw indications for lower induced absorption if a lower density of charge-carriers was created, thus absorption due to IVB should be the dominant process here. 
A comparison between experimental and first-principles data for $\Delta(\varepsilon_2)$ below 3\,eV in Fig.\,\ref{fig:transients}\,(d) (inset) shows good agreement in the line shape of measured and calculated spectral features.
The sub-gap energy-range between 2 and 3\,eV is dominated by contributions from IVB transitions that become allowed in the presence of free holes.
The computational results do not account for phonon-assisted processes, which likely explains why the computational data underestimates the experiment at these energies. 
Conduction-conduction band transitions do not significantly contribute in this energy range. 
The appearance of the low-energy absorption indicates that the spectral weight of absorption is transferred from the fundamental absorption edge to lower energies because the total number of charge carriers remains constant, which is known as the sum rule \cite{Franta2013}. 

\section{Conclusion}

The development of fs-time-resolved spectroscopic ellipsometry allows unambiguous determination of the complex, fre\-quen\-cy-dependent dielectric function with sub-ps temporal resolution in a wide spectral range. It is hence a unique tool to study the dynamics of electronic systems in solids that beats conventional transient spectroscopy in several areas. 
Investigating a UV-pumped ZnO thin film, we were able to discriminate different processes of the non-equilibrium charge-carrier dynamics of this strongly photo-degenerate semiconductor. 
In contrast to previous experiments, we are able to largely minimize risks for mis-interpretations that can originate from mixing effects of changed absorption and refractive index. 
We observe partial blocking and screening of near-band-edge and exciton absorption due to occupation of the electronic states. 
A non-vanishing excitonic absorption enhancement hints at the occurrence of Mahan excitons. 
Furthermore, intra-valence-band transitions become possible when holes scatter to the edges of the Brillouin zone, 
their fast response time renders them interesting for optoelectronic switching devices. 
Finally, there is evidence for hot-phonon effects, from both a delayed charge-carrier relaxation and an increase in the exciton-phonon-complex energy. 
The described dynamics are crucially dependent on the pump wavelength and hence the excess energy obtained by the carriers which determines their effective temperature.
From our data we can also conclude that the high density of hot charge carriers does not trigger the Mott transition. 
The survival of the excitonic absorption reflects directly the non-equilibrium distribution of the highly excited charge carriers. 
These results stimulate a demand for the development of new theories describing high-density exciton systems beyond the present state.

\section*{Acknowledgement}

We thank Peter Schlupp for growing the thin film and Michael Lorenz (both Universit{\"a}t Leipzig) for X-ray diffraction measurements. 
We gratefully acknowledge valuable discussions with Christoph Cobet, Martin Feneberg, Daniel Franta, Kurt Hingerl, Michael Lorke, Bernd Rheinl\"ander, Chris Sturm and Marcel Wille, and lab support by Martin P\v{r}e\v{c}ek. 
And we thank Alina Pranovich for graphics support. 
Parts of this work have been funded by the Deutsche Forschungsgemeinschaft (DFG, German Research Foundation), SFB 762 - Projektnr. 31047526 (project B03), and FOR 1616 (SCHM2710/2). 
O.H. acknowledges the Leipzig School of Natural Sciences BuildMoNa. 
Experimental development at ELI Beamlines was funded by the project "Advanced research using high intensity laser produced photons and particles" (ADONIS), Reg. n. CZ.02.1.01/0.0/0.0/16\_019/0000789, from the European Regional Development Fund, and the National Program of Sustainability II project “ELI Beamlines - International Center of Excellence” (ELISus), project code: LQ1606. 
S.E. was partially supported by the project “Structural dynamics of biomolecular systems” (ELIBIO), reg. no. CZ.02.1.01/0.0/0.0/15\_003/0000447, from the European Regional Development Fund. 
J.A. acknowledges support of the Ministry of Education, Youth and Sports as part of targeted support from the National Programme of Sustainbility II and the Chalmers Area of Advanced Materials Science. 
J.A.L. and A.S. were supported by the National Science Foundation under Grant Nos. DMR-1555153 and CBET-1437230, and computer time was provided as part of the Blue Waters sustained-petascale computing project, which is supported by the National Science Foundation (awards OCI-0725070 and ACI-1238993) and the state of Illinois. 
S.Z. was supported by the National Science Foundation, Grant No. DMR-1505172.

\section*{References}
\bibliographystyle{unsrt}
\bibliography{references}

\clearpage


\section*{Supplementary material}

# Supplementary material:
# Ultrafast dynamics of hot charge carriers in an oxide semiconductor probed by femtosecond spectroscopic ellipsometry

Steffen Richter[1,2,*], Oliver Herrfurth[2], Shirly Espinoza[1], Mateusz Rębarz[1], Miroslav Kloz[1], Joshua A. Leveillee[3], André Schleife[3], Stefan Zollner[4,5], Marius Grundmann[2], Jakob Andreasson[1], Rüdiger Schmidt-Grund[2,6]

[1] ELI Beamlines/Fyzikální ústav AV ČR, v.v.i., Za Radnicí 835, 25241 Dolní Břežany, Czech Republic
[2] Universität Leipzig, Felix-Bloch-Institut für Festkörperphysik, Linnéstr. 5, 04103 Leipzig, Germany
[3] University of Illinois, Dep. of Materials Science and Engineering, 1304 W. Green Str., Urbana, IL 61801, USA
[4] New Mexico State University, Department of Physics, PO Box 30001, Las Cruces, NM, 88003-8001, USA
[5] Fyzikální ústav AV ČR, v.v.i., Sekce optiky, Na Slovance 2, 18221 Praha, Czech Republic
[6] Technische Universität Ilmenau, Institut für Physik, Weimarer Str. 32, 98693 Ilmenau, Germany
*Present address: Linköpings universitet, Institutionen för fysik, kemi och biologi, 58183 Linköping, Sweden.

May 2020

## I. Measurement scheme and data reduction

In contrast to sapphire, calcium-fluoride-based white light generation offers more UV intensity up to $3.6\,\text{eV}$, but the crystal needs to be moved during creation of continuum white light in order to protect the crystal from heat damage. This movement and CCD warm-up yield fluctuating intensity spectra. The situation is very different from any other ellipsometer where the light source is stable at least over the time of a complete revolution of the rotating element. We circumvent the problem by applying a two-chopper scheme as depicted in Fig. S1. Repeatedly, four different intensity signals, "pump & probe" (S1), "pump only" (S2), "probe only" (S3), and "dark" (S4) are measured. Hence, at any time, background- or even luminescence-corrected "pump & probe" ($R_j^p(E) \equiv I_{S1} - I_{S2}$) as well as "probe only" ($R_j^0(E) \equiv I_{S3} - I_{S4}$) spectra are obtained for each compensator angle $\alpha_j$. However, they are still subject to intensity fluctuations as can be seen in Fig. S2. Using all four phases created by the chopper, we can obtain proper reflectance difference spectra

$$\left(\frac{\Delta R(E)}{R(E)}\right)_j = \frac{R_j^p(E) - R_j^0(E)}{R_j^0(E)} \equiv \frac{I_{S1} - I_{S2}}{I_{S3} - I_{S4}} - 1\,. \tag{S1}$$

Here, $j = 1\ldots10$ indicates a certain compensator azimuth angle during the measurement. In the experiments,

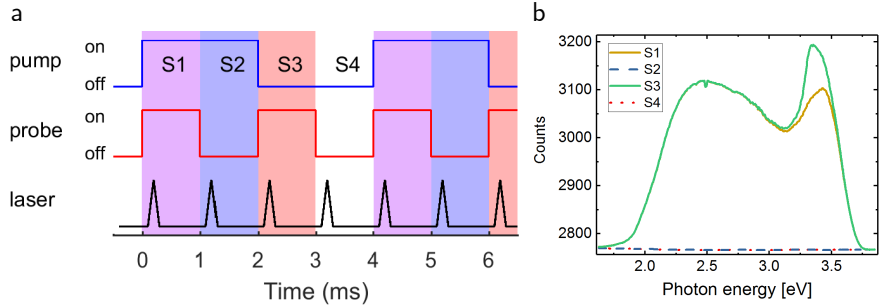

**Figure S1: a** Visualization of the two-chopper scheme. **b** Example of a set of measured intensity spectra at $\Delta t = 400\,\text{fs}$ and compensator azimuth angle $100°$.

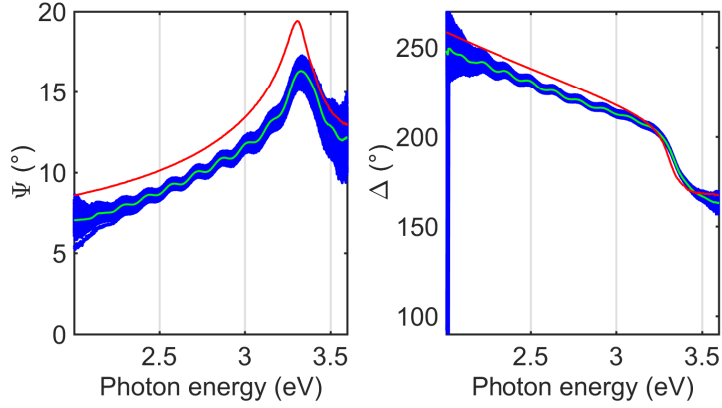

**Figure S2:** Spectra of the ellipsometric parameters $\Psi, \Delta$ obtained from all "probe only" reflectance measurements ($R^0$, evaluating only chopper phases S3 and S4) obtained during a delay line scan. The green line indicates the average and the red line shows reference spectra obtained with a commercial ellipsometer. Note that these spectra are only shown as a benchmark. They are prone to offsets and modulations arising from long-term changes in the whitelight spectra or intensities. The oscillatory shape of those errors originates from the quarterwave plate. For the time-resolved ellipsometry, the reflectance difference signal is evaluated instead. These are robust against long-term changes.

the compensator was rotated in 10 steps of $50°$. The polarizer was set at $-45°$, the analyzer at $+45°$. Each spectrum was averaged over 500 pulses.

To compute the ellipsometric angles we utilize Moore-Penrose pseudo-inversion (ordinary least-squares regression) in a Müller matrix formalism for each photon energy and delay time [S1]: The Müller matrix of the sample in isotropic or pseudo-isotropic configuration is given as

$$M_{\text{sample}} = \begin{pmatrix} M_{11} & M_{12} & 0 & 0 \\ M_{12} & M_{11} & 0 & 0 \\ 0 & 0 & M_{33} & M_{34} \\ 0 & 0 & -M_{34} & M_{33} \end{pmatrix} \tag{S2}$$

$$= M_{11} \begin{pmatrix} 1 & -N & 0 & 0 \\ -N & 1 & 0 & 0 \\ 0 & 0 & C & S \\ 0 & 0 & -S & C \end{pmatrix} = M_{11} \begin{pmatrix} 1 & -\cos(2\Psi) & 0 & 0 \\ -\cos(2\Psi) & 1 & 0 & 0 \\ 0 & 0 & \sin(2\Psi)\cos(\Delta) & \sin(2\Psi)\sin(\Delta) \\ 0 & 0 & -\sin(2\Psi)\sin(\Delta) & \sin(2\Psi)\cos(\Delta) \end{pmatrix}.$$

For each compensator angle $\alpha_j$, the Müller matrix $M^{\text{det},j}$ shall represent a respectively oriented compensator followed by an polarizer (analyzer) as in the experiment. Likewise, $M^{\text{prep}}$ shall represent the Müller matrix of a polarizer at the angle of the polarizer in the experiment. Having measured $N$ different configurations (compensator angles) $j = 1 \ldots N$, we can introduce a $4 \times N$ setup coefficient matrix $\hat{M}_{\text{setup}}$. Its $j$th column can be written as

$$\hat{M}_{\text{setup}}^j = \begin{pmatrix} M_{11}^{\text{det},j} M_{11}^{\text{prep}} + M_{12}^{\text{det},j} M_{21}^{\text{prep}} \\ -M_{11}^{\text{det},j} M_{21}^{\text{prep}} - M_{12}^{\text{det},j} M_{11}^{\text{prep}} \\ M_{13}^{\text{det},j} M_{31}^{\text{prep}} + M_{14}^{\text{det},j} M_{41}^{\text{prep}} \\ M_{13}^{\text{det},j} M_{41}^{\text{prep}} - M_{14}^{\text{det},j} M_{31}^{\text{prep}} \end{pmatrix}. \tag{S3}$$

With the row vector $\vec{R}$ containing the $N$ intensity values $R_j$ for each compensator angle $\alpha_j$, it holds

$$M_{11} (1, N, C, S) = \vec{R} \hat{M}_{\text{setup}}^{\text{T}} (\hat{M}_{\text{setup}} \hat{M}_{\text{setup}}^{\text{T}})^{-1}. \tag{S4}$$

Instead of using the "pump & probe" intensity spectra $R_j^{\text{p}}(E)$, the reflectance difference signal $(\Delta R(E)/R(E))_j$ is applied to ideal (theoretical) intensity spectra of the unexcited sample $R_j^{00}(E)$ which are computed from reference ellipsometry spectra. Hence, $R_j = R_j^{00}(1 + (\Delta R/R)_j)$.

In a final step, the Müller matrix elements can be transferred to ellipsometric angles and the degree of polarization ($DOP$):

$$\Psi = \frac{1}{2} \tan^{-1} \left( \frac{\sqrt{C^2 + S^2}}{N} \right), \tag{S5}$$

$$\Delta = \tan^{-1} \left( \frac{S}{C} \right), \tag{S6}$$

$$DOP = \sqrt{N^2 + C^2 + S^2}, \tag{S7}$$

requiring $\Psi \in [0°, 90°]$ and $\Delta \in (90°, 270°)$ if $C < 0$, $\Delta \in (0°, 90°) \cup (270°, 360°)$ if $C > 0$. It should be noted that $\Psi$ and $\Delta$ are to first order unaffected by depolarization, i.e., the above equations intrinsically involve only the non-depolarizing part of the Müller matrix. Depolarization results in $M_{22} \neq M_{11} = 1$ in contrast to Eq. S2. However, as in the experimental configuration the input polarization was chosen to be linear at azimuth angle $\pm 45°$, $M_{22}$ is not probed and thus depolarization does not affect the data reduction. The non-depolarizing Mueller matrix is obtained by replacing $(N, C, S)$ by $(N, C, S)/DOP$.

Finally, the obtained data reveal an imprinted chirp of the white light, i.e. propagation through the $CaF_2$ window and the support of the wiregrid polarizers caused light of longer wavelength to arrive earlier at the sample than light of shorter wavelength. This is illustrated in Fig. S3. An even polynomial function is used to describe this chirp and adjust the zero delay for each photon energy. Data is interpolated accordingly.

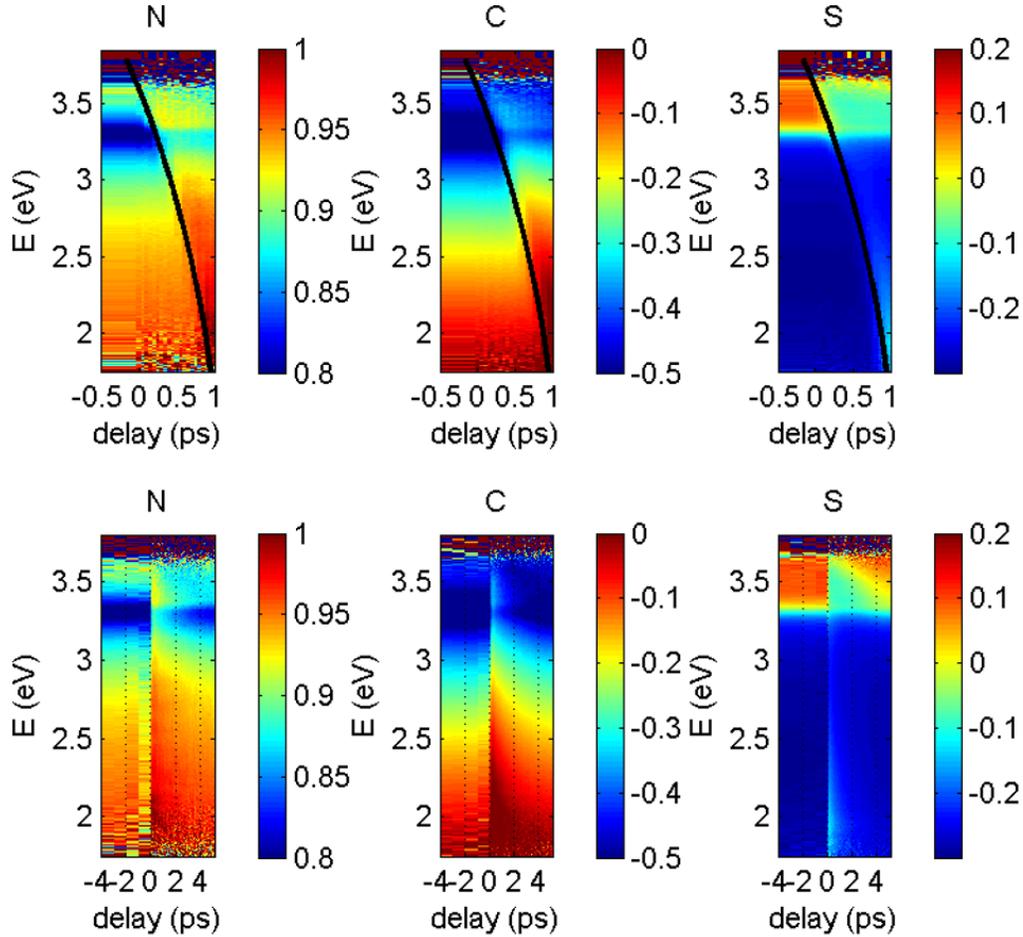

**Figure S3:** Experimentally obtained Müller matrix elements $N$, $C$, $S$ during the first picoseconds. Top row: Data as obtained from the experiment with clear indication of the chirped whitelight pulse. Black curves show the polynomial function used to describe the true delay zero. Bottom row: Data after chirp correction by adjusting the zero positions for each photon energy.

As described in the article, the obtained Müller matrix spectra $N(E)$, $C(E)$, $S(E)$ were modeled by means of transfer matrix calculations based on a model for a 30 nm ZnO film on fused silica substrate. While the optical constants (DF) of the substrate are kept constant along all delay times, the DF of the ZnO film is individually described by a B-spline function which is fitted to match the experimental Müller matrix spectra. Example spectra are shown in Fig. S4.

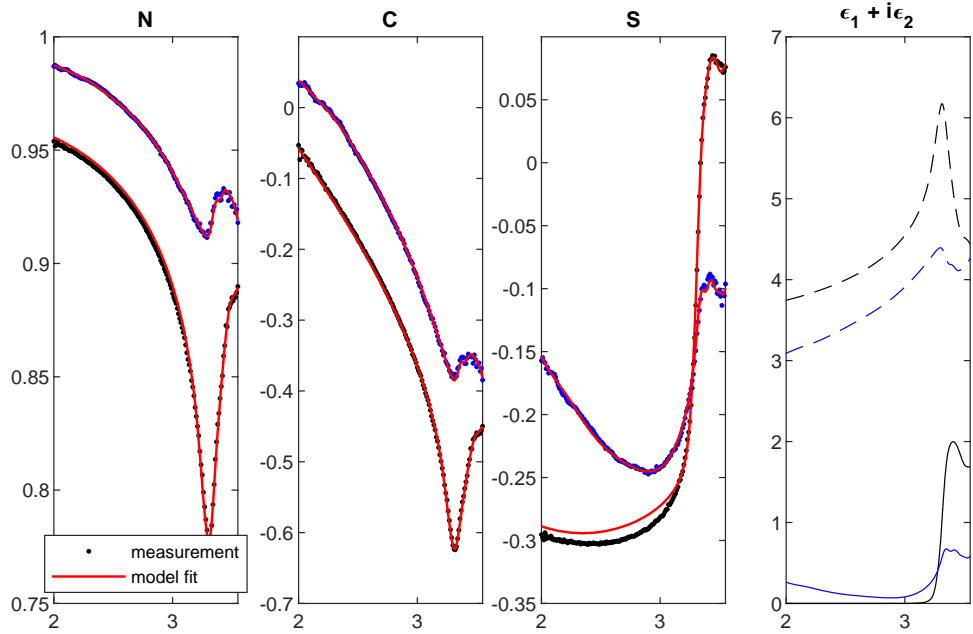

**Figure S4:** Example modeling of the obtained Müller matrix spectra $N(E)$, $C(E)$, $S(E)$ for pump-probe delay -7 ps (black) and 200 fs (blue). The respective B-Spline DFs are shown at the right panel ($\varepsilon_1$ dashed, $\varepsilon_2$ solid lines).

## II. Error estimation

In order to estimate the errors of our ellipsometric parameters and obtained DF, different error sources need to be considered. For the transient experiment, the most critical limitation in order to observe pump-induced changes is the quality of the reflectance-difference raw signal. In order to estimate the error in $\Delta R/R$, we show the measured data at negative delays in Fig. S5. On average, the deviation is smaller than $\pm 0.003$ between 2 and 3.4 eV. However, it is clear that the data quality drops for large photon energies where less or no whitelight is generated (3.6 eV marking approximately the maximum) as well as for lower photon energies where a filter blocked the generating laser at 1.55 eV (1.8 eV being the current minimum energy for acceptable data).

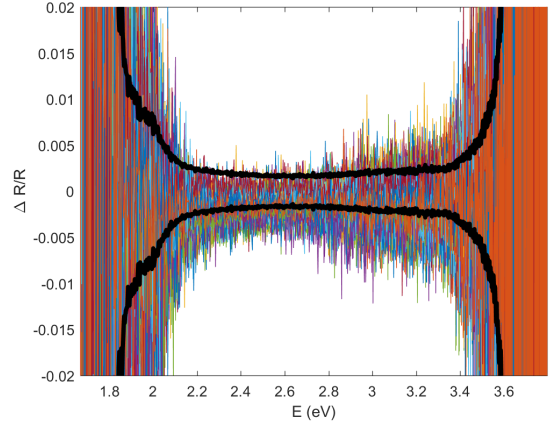

**Figure S5:** Uncertainty of the measured $\Delta R/R$ signal for the first 10 delays set in the experiment. All 100 spectra (all 10 compensator angles) correspond to negative delays and were obtained more than 500 fs before the pump pulse arrived. Thick black lines illustrate $\pm$ the mean of the absolute values in order estimate the average maximum deviation from 0.

The obtained error bars for $\Delta R/R$ (thick black lines in Fig. S5) are the basis for Monte-Carlo simulations we performed in order to estimate the largest possible error in $N$, $C$, $S$, $\Psi$, $\Delta$. Those ellipsometric parameters can also be represented as so called pseudo dielectric-function $< \varepsilon >$ which would correspond to the material's DF in case of a semi-infinite, isotropic sample with perfect surface. It holds

$$< \varepsilon > = \sin^2(AOI) \left( 1 + \left( \tan(AOI) \frac{1-\rho}{1+\rho} \right)^2 \right), \tag{S8}$$

where $AOI$ is angle of incidence and $\rho = \tan(\Psi) e^{-i\Delta}$. The estimated error of $< \varepsilon >$ can hold as a first approximation for the error of the DF that we obtain for our film by modeling with the B-spline function.

Monte-Carlo simulations of our data reduction accounting for the uncertainties in $\Delta R/R$ result in statistical errors for the ellipsometric parameters shown in Fig. S6. The shown standard deviations mark the limits of our experimental setup in terms of sensitivity to pump-probe induced changes.

As a maximum-possible-error estimation of the systematic error, we performed the same Monte-Carlo simulations with assumed uncertainties for the azimuth angles of polarizers and compensator, as well as the retardance of the compensator. All those errors can arise from slight mis-alignment. Additionally, the used reference ellipsometry data might have an error which we did not consider here. The results are shown in Fig. S7 where the input uncertainties are given in the caption. As mentioned, this is an estimate for the largest possible error that might result from experimental imperfections. Especially the assumed max. 10% uncertainty of the compensator retardance are a *worst-case* scenario.

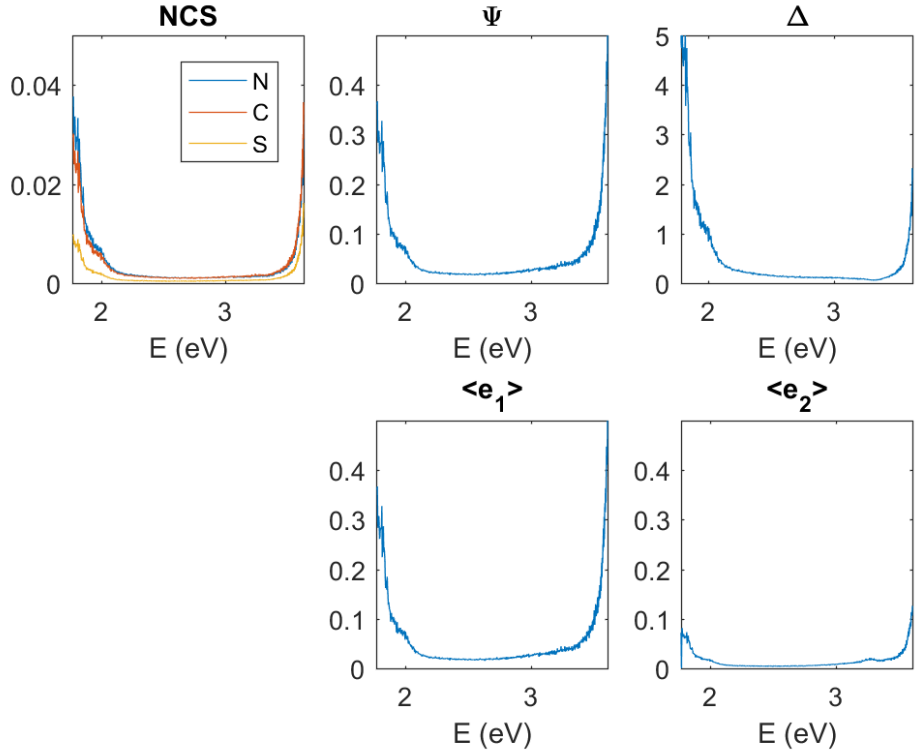

**Figure S6:** Statistical uncertainty arising from the uncertainty of $\Delta R/R$ as standard deviation obtained by Monte-Carlo simulations. These errors define the quality of the experiment.

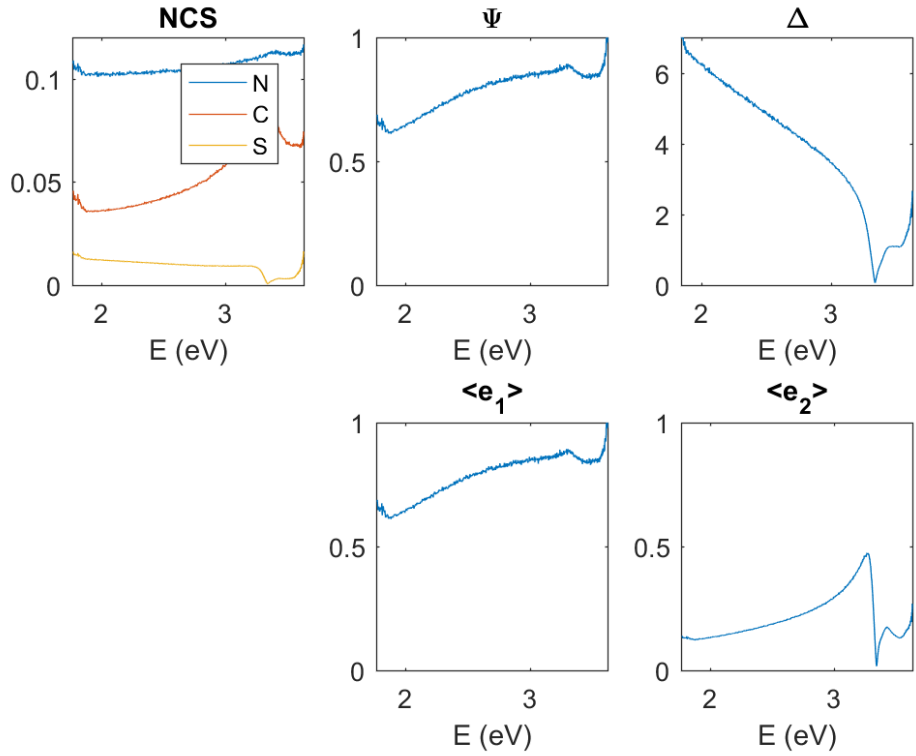

**Figure S7:** Maximum total error given as standard deviation calculated by taking into account the uncertainty of $\Delta R/R$, an angle uncertainty of $0.5°$ for the polarizer, $1°$ for the analyzer and $0.55°$ for the compensator, as well as 10% uncertainty of the compensator retardance. Statistics obtained by means of Monte-Carlo simulations.

## III. General sample characterization

Time-resolved photoluminescence (PL) spectroscopy conducted with a streak camera reveals information on the temporal evolution of the occupation of electronic states. The sample was optically excited with 4.67 eV pulses of a frequency-tripled Ti:Sapphire laser (3 MHz, 150 fs, 1 nJ). Figure S8 a shows the transient photoluminescence at the absorption edge of ZnO (3.28 eV), which is much less intense compared to the defect luminescence centered at 2.4 eV. This hints at the defect-rich crystal growth induced by the amorphous $SiO_2$ substrate. The ratio of near-band-edge to defect-related luminescence is not constant over the sample surface.

We model the transient UV-PL (Fig. S8 b, c) with onset $\tau_o$ and decay time $\tau_d$ of roughly 4 ps, which we expect to be limited by the time resolution of our streak camera. The preferred radiative recombination channel appears to be related to defect states having an order of magnitude higher onset $\tau_o = 60$ ps as well as decay times $\tau_{d,1} = 80$ ps and $\tau_{d,2} = 415$ ps. These time constants match the late absorption recovery that is observed in the time-resolved ellipsometry experiment. The excited electron population seems to be not yet fully recombined after 2 ns, corresponding to the time scale for vanished band bending observed in the time-resolved spectroscopic ellipsometry data.

The X-ray data (Fig. S9) confirm $c$-plane orientation of the thin film and show the response of the amorphous substrate. The FWHM of the ZnO (002) rocking curve is larger compared to other PLD-grown ZnO thin films [S2]. The grain size is estimated to be on the order of the film thickness using the Scherrer formula.

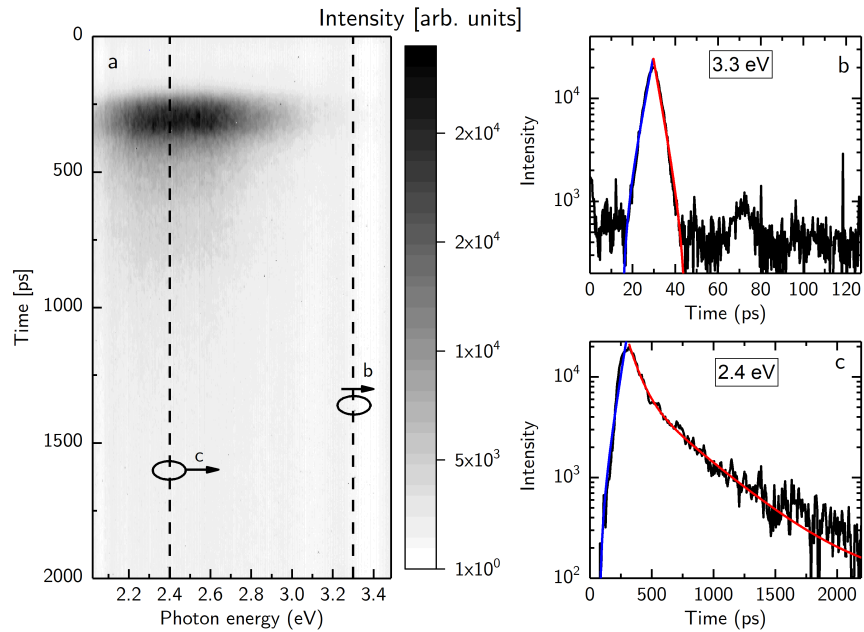

**Figure S8: a** Time-resolved photoluminescence measured by a streak camera. The dashed lines indicate the transients shown in panel **b, c**. Blue (red) lines indicate an exponential model fit to obtain characteristic onset (decay) times.

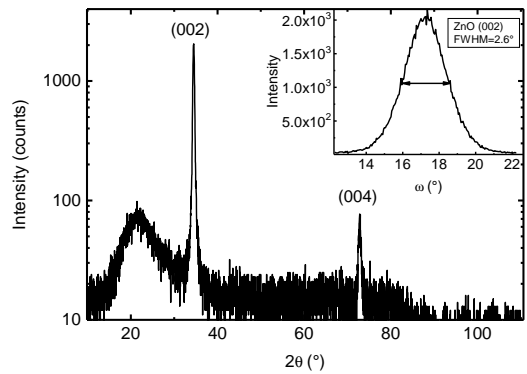

**Figure S9:** $2\theta$-$\omega$ scan of the 30 nm thick ZnO film on an $SiO_2$ substrate. The inset shows the rocking curve of the ZnO (002) peak.

# IV. Optical transitions in ZnO

With the symmetry assignments of the bands according to [S3], the dipole-allowed transitions for the electric field oriented perpendicular to the optic axis ($\mathcal{E} \perp c$) in wurtzite ZnO (space group 186) are listed in table IV; cf. also [S4, 5]. Only relevant bands at high-symmetry points of the Brillouin zone are considered and Koster notation of the irreducible representations is used. Transitions for $\mathcal{E} \parallel c$ are only allowed between states of the same symmetry representation.

| direction | point group symmetry | dipole operator representation | allowed transitions for $\mathcal{E} \perp c$ |
|-----------|----------------------|-------------------------------|-----------------------------------------------|
| $\Gamma$, $\Delta$, A | 6mm ($C_{6v}$) | $\Gamma_5$ | $\Gamma_1 \leftrightarrow \Gamma_5$, $\Gamma_2 \leftrightarrow \Gamma_5$ |
| | | | $\Gamma_3 \leftrightarrow \Gamma_6$, $\Gamma_5 \leftrightarrow \Gamma_6$ |
| P, K, H | 3mm ($C_{3v}$) | $\Gamma_3$ | $\Gamma_1 \leftrightarrow \Gamma_3$, $\Gamma_2 \leftrightarrow \Gamma_3$ |
| | | | $\Gamma_3 \leftrightarrow \Gamma_3$ |
| U, M, L | 2mm ($C_{2v}$) | $\Gamma_3$ | $\Gamma_1 \leftrightarrow \Gamma_3$, $\Gamma_2 \leftrightarrow \Gamma_4$ |

For the reciprocal-space directions corresponding to monoclinic $C_s/C_{1h}$ symmetry (R, $\Sigma$ as ..m, and S, T as .m.), where the $c$-direction of the crystal is parallel to the respective mirror planes, the assignment of band symmetries and transitions is generally more complex. The dipole operator would transform generally like $\Gamma_1$, in some cases like $\Gamma_3$.

# V. Charge-carrier density

Assuming linear absorption, the density $N$ of photo-excited electron-hole pairs in the film can be estimated as

$$ N \approx E_{\text{pulse}} \frac{\lambda_{\text{pump}}}{hc_0}(1-R) \left[ 1 - \exp\left( -\frac{\alpha_{\text{film}}d_{\text{film}}}{\cos(\theta_{\text{film}})} \right) \right] \left[ \frac{\cos(\theta_{\text{pump}})}{(d_{\text{pump}}/2)^2 \pi} \frac{\cos(\theta_{\text{film}})}{d_{\text{film}}} \right] $$

with

| quantity | meaning |
|----------|---------|
| $E_{\text{pulse}} = 1\,\mu\text{J}$ | pump pulse energy |
| $\lambda_{\text{pump}} = 266\,\text{nm}$ | pump photon wavelength |
| $d_{\text{pump}} = 400\,\mu\text{m}$ | pump spot diameter |
| $\theta_{\text{pump}} = 40°$ | pump incidence angle |
| $\theta_{\text{film}} = 19°$ | pump angle in the film (with refr. index $n \approx 2$) |
| $d_{\text{film}} = 30\,\text{nm}$ | ZnO film thickness |
| $\alpha_{\text{film}} = (50\,\text{nm})^{-1}$ | ZnO absorption coefficient |
| $R = 0.2$ | surface reflectance |

This formula accounts for reflectance losses and an effectively enlarged pump spot as well as film thickness at oblique incidence. It does not account for reflectance from the film-substrate interface which increases the absorption (in fact, here it would increase the intensity available for absorption by about 1%). With the experimental parameters above, the effective energy density of the pump was about $500\,\mu\text{J/cm}^2$, already taking into account 20% reflection losses. With a penetration depth of $50\,\text{nm}$ in ZnO, roughly 45% of the pump power is absorbed in the film. The substrate is transparent for light of $266\,\text{nm}$ wavelength. Furthermore, only about 87% of the entire pulse energy are contained within the $1/e^2$ area which defines $d_{\text{pump}}$. However, the latter is compensated by the non-even beam profile as we probe only the central $200\,\mu\text{m}$ of the $400\,\mu\text{m}$ diameter of the excited area. With the numbers above given, one arrives at $N \approx 9.75 \times 10^{19}\,\text{cm}^{-3}$.

It should be noted that we assume linear absorption. In fact, absorption bleaching of the material can also take place at the laser energy if the corresponding initial and final states are already empty or filled, respectively. This effect can only matter if the excitation pulse is sufficiently short so that carrier scattering cannot compensate for the bleaching during the time of the excitation pulse. In other words, there is a limit for the highest achievable density of excited electron-hole pairs for ultrashort laser pulses. Even with higher pump power, parts of that laser pulse would not be absorbed. This could be an explanation why the excitonic absorption peaks do not completely vanish, meaning the excitation density is overestimated. However, the estimated number of excited electron-hole pairs in the experiment here seems to be consistent with other works, using different pulsed laser sources. Finally, there are preliminary indications that shorter laser pulses in the order of $20\,\text{fs}$ instead of $35\,\text{fs}$ induce less IVB absorption. This hints at absorption bleaching.

## VI. Charge-carrier statistics

Upon optical pumping with a 266 nm ($E_{\text{pump}} = 4.66\,\text{eV}$) laser pulse, the excited electrons and holes obtain different amounts of excess energy, related to their effective masses (parabolic band approximation) [S6]:

$$\Delta E_{\text{e}} = \frac{E_{\text{pump}} - E_{\text{gap}}}{1 + m_{\text{e}}/m_{\text{h}}}, \qquad \Delta E_{\text{h}} = \frac{E_{\text{pump}} - E_{\text{gap}}}{1 + m_{\text{h}}/m_{\text{e}}}$$

With a bandgap energy of $E_{\text{gap}} \approx 3.4\,\text{eV}$, electron effective mass $m_{\text{e}} = 0.24 m_0$ [S7] and hole effective mass $m_{\text{h}} = 0.59 m_0$ [S8] ($m_0$ being the free electron mass), it follows $\Delta E_{\text{e}} \approx 0.90\,\text{eV}$ and $\Delta E_{\text{h}} \approx 0.36\,\text{eV}$.

Assuming the free-electron/hole gas as an ideal gas, an average kinetic energy corresponding to the excess energy $\Delta E_{\text{e/h}}$ is related to an effective temperature $T_{\text{e/h}}$ by

$$\Delta E_{\text{e/h}} = \frac{3}{2} k_{\text{B}} T_{\text{e/h}}$$

with Boltzmann factor $k_{\text{B}}$. From this, we can estimate initial effective temperatures for the charge carriers as $T_{\text{e}} \approx 7000\,\text{K}$ and $T_{\text{h}} \approx 2800\,\text{K}$.

While the effective charge-carrier temperatures express directly the average excess energy of excited electrons and holes, their density $N_{\text{e}} = N_{\text{h}}$ is given as [S9]:

$$N_{\text{e}} = N_{\text{C}} \frac{2}{\pi} F_{1/2} \left( \frac{E_{\text{F}}^{\text{e}} - E_{\text{CB}}}{k_{\text{B}} T_{\text{e}}} \right), \qquad N_{\text{h}} = N_{\text{V}} \frac{2}{\pi} F_{1/2} \left( \frac{E_{\text{VB}} - E_{\text{F}}^{\text{h}}}{k_{\text{B}} T_{\text{h}}} \right)$$

with the Fermi-Dirac integral $F_{1/2}$. $E_{\text{CB/V}}$ are the energies of the conduction-band minimum and valence-band maximum, respectively. The effective densities of states (DOS) at the conduction band minimum and valence band maximum are, respectively

$$N_{\text{C}} = 2 \left( \frac{m_{\text{e}} k_{\text{B}} T_{\text{e}}}{2\pi \hbar^2} \right)^{3/2}, \qquad N_{\text{V}} = 2 \left( \frac{m_{\text{h}} k_{\text{B}} T_{\text{h}}}{2\pi \hbar^2} \right)^{3/2}$$

It can be estimated that $N_{\text{C}}(T_{\text{e}} \approx 7000\text{K}) \approx 3.3 \cdot 10^{20}\,\text{cm}^{-3}$ and $N_{\text{V}}(T_{\text{h}} \approx 2800\text{K}) \approx 3.2 \cdot 10^{20}\,\text{cm}^{-3}$ for the estimated carrier temperatures [1]. However, it should be noted that the temperature dependence of those effective DOS's results only from a substitution of the integrating variable from $E$ to $E/k_{\text{B}} T_{\text{e/h}}$ when expressing

$$N_{\text{e/h}} = \int \text{DOS}(E)/(1 + e^{(E - E_{\text{F}}^{\text{e/h}})/k_{\text{B}} T_{\text{e/h}}}) \, \text{d}E \tag{S9}$$

---

[1] At room temperature, $N_{\text{C}} \approx 3 \cdot 10^{18}\,\text{cm}^{-3}$ and $N_{\text{V}} \approx 1 \cdot 10^{19}\,\text{cm}^{-3}$

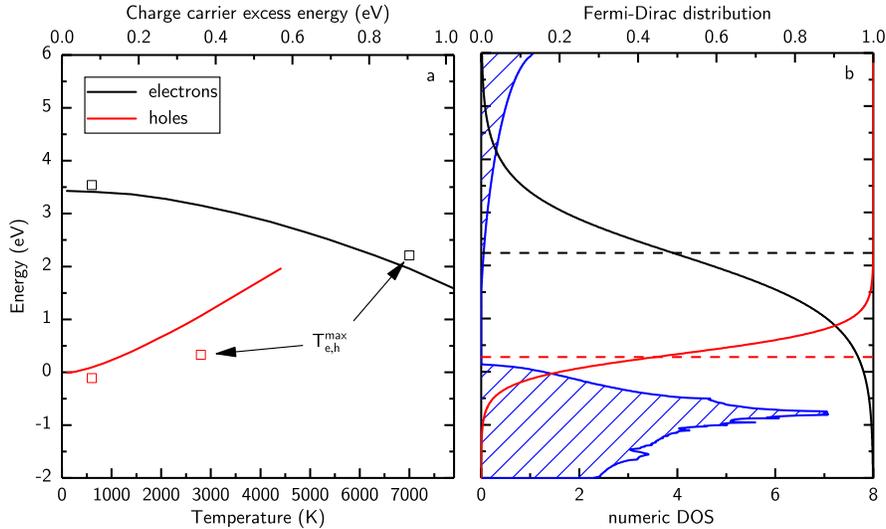

**Figure S10: Quasi Fermi-energies and distributions at high charge-carrier temperature: a:** quasi Fermi-energies for electrons (black) and holes (red) depending on the carrier temperature for a fixed carrier density of $10^{20}\,\text{cm}^{-3}$. Lines represent theoretical results which are computed by evaluating the Fermi integral for electrons and holes using the ground-state density of states (DOS) computed within density functional theory (see **b**). Symbols represent the evaluation as discussed in the text here with assumed non-parabolicity effect. **b:** First-principles numeric DOS (blue) and Fermi-Dirac distribution functions (red/black solid lines) for the situation with $T_{\text{e}} = 7000\,\text{K}$ and $T_{\text{h}} = 2800\,\text{K}$. Dashed lines highlight the quasi Fermi-energies.

through $F_{1/2}$ as above. When estimating quasi Fermi-energies[2] $E_F^{e/h}$ for the hot charge carriers, it is important to understand both, their dependence on carrier density and temperature. Zero-temperature approximations as introduced in Eq. 5 in the article to account for excitonic corrections do not hold any more. At a given temperature, a higher carrier density will clearly shift the quasi Fermi-energies towards/into the respective bands, i.e. $E_F^e$ increases and $E_F^h$ decreases. However, the effect of high temperatures (at a given carrier density) is more sophisticated: Evaluating the Fermi-Dirac integral with constant prefactors $N_{C/V}$ shows that the quasi Fermi-energies would shift further towards/into the bands if the effective temperatures are higher. On the other hand the temperature dependence of $N_{C/V}$ yields exactly the opposite and is even more dominant. Thus, in total, despite the high density of charge carriers, the quasi Fermi-energies are pushed into the bandgap due to the high carrier temperatures. Fitting the Fermi-Dirac integral to the initial density $N_{e/h} \approx 10^{20}\,\mathrm{cm}^{-3}$ results in estimates on the order of $E_F^e - E_{CB} \approx$ -660 meV and $E_{VB} - E_F^h \approx$ -260 meV for the above-obtained effective temperatures. This means that both quasi Fermi-energies are within the bandgap, which is consistent with the numerical first-principles computations, see Fig. S10. Compared with the intrinsic Fermi energy $E_F$ which is typically close to the conduction-band minimum due to intrinsic free electrons, $E_F^e$ is shifted even further into the bandgap.

It should be noted that those estimates rely on parabolic approximations. The non-parabolicity of the bands yields another strong increase of the DOS through increasing effective masses for energies far from the minimum of the conduction and maximum of the valence band. A doubled effective mass causes the distances of the quasi Fermi-levels to the valence/conduction band maximum/minimum to increase to roughly twice the calculated values. For the conduction band with the obtained carrier temperature $T_e$ we can estimate from a non-parabolicity parameter on the order of $0.4\,\mathrm{eV}^{-1}$ [S10] that $E_F^e - E_{CB}$ should be in the order of 1 eV below the conduction band minimum [S11]. Assuming a similar non-parabolicity for the valence band results consequently in $E_{VB} - E_F^h \approx -300$ meV.

## VII. Charge-carrier screening

It is interesting to estimate the plasma frequencies related to the electrons and holes. With the above-mentioned parabolic masses, we obtain for the unscreened plasma frequencies for $N_e = N_h = 10^{20}\mathrm{cm}^{-3}$

$$\omega_{p0}^e = \sqrt{\frac{N_e e^2}{\varepsilon_0 m_e}} \approx 1150\,\mathrm{THz}\,, \qquad \omega_{p0}^h = \sqrt{\frac{N_h e^2}{\varepsilon_0 m_h}} \approx 730\,\mathrm{THz}\,.$$

These values correspond to 0.76 eV and 0.48 eV for electrons and holes, respectively, and are hence much larger than the LO phonon energies of about 72 meV. Screening occurs on the time scale of a period of plasma oscillations, hence $2\pi/\omega_{p0} \approx$ 5.5 fs and 8.6 fs for electrons and holes, respectively.

Assuming $\varepsilon_\infty \approx 4$, the screened plasma frequencies are

$$\omega_p^e = \sqrt{\frac{N_e e^2}{\varepsilon_0 \varepsilon_\infty m_e}} \approx 580\,\mathrm{THz}\,, \qquad \omega_p^h = \sqrt{\frac{N_h e^2}{\varepsilon_0 \varepsilon_\infty m_h}} \approx 370\,\mathrm{THz}\,.$$

These values correspond to to 0.38 eV and 0.24 eV, which one plasma oscillation period being 11 fs or 17 fs. Note that $\varepsilon_\infty$ decreases immediately after pumping.

Finally, taking into account that electrons and holes are not independent of each other when it comes to plasma oscillations, $m_e$ and $m_h$ have to be replaced by the effective excitonic mass $m_{eh} = 1/(m_e^{-1} + m_h^{-1}) \approx 0.17 m_0$. We obtain effective unscreened and screened plasma frequencies as follows:

$$\omega_{p0}^{eh} = \sqrt{\frac{N e^2}{\varepsilon_0 m_{eh}}} \approx 1370\,\mathrm{THz}\,,$$

$$\omega_p^{eh} = \sqrt{\frac{N e^2}{\varepsilon_0 \varepsilon_\infty m_{eh}}} \approx 680\,\mathrm{THz}\,.$$

Respective plasma energies and period times are 0.45 eV (0.90 eV) and 9.2 fs (4.6 fs) respectively for the screened (unscreened) definitions. In any case, screening occurs instantaneous within the temporal resolution of our experiment.

Similarly, we can estimate the Debye screening lengths for the hot charge carriers:

$$\lambda_D^e = \sqrt{\frac{\varepsilon_0 k_B T_e}{N_e e^2}} \approx 5.8\,\mathrm{\mathring{A}}\,, \qquad \lambda_D^h = \sqrt{\frac{\varepsilon_0 k_B T_h}{N_h e^2}} \approx 3.7\,\mathrm{\mathring{A}}\,.$$

---

[2] The term Fermi energy or Fermi level is used in consistency with most literature on semiconductors. However, precisely spoken, we refer actually to the chemical potential and note that Fermi energy is the limit of the chemical potential at zero temperature.

Hence, the high density of the charge carriers reduces the screening to a length scale corresponding to the unit cell of the ZnO crystal. The effective screening length by electrons and holes acting at the same time can be estimated according to Versteegh *et al.* [S12] to

$$\lambda_{\mathrm{D}} = \left( \frac{1}{\lambda_{\mathrm{D}}^{\mathrm{e}\,2}} + \frac{1}{\lambda_{\mathrm{D}}^{\mathrm{h}\,2}} \right)^{-1/2} \approx 3.1\,\text{Å} \,.$$

This is to be compared to the exciton Bohr radius $a_{\mathrm{B}}$ which is about $2\,\mathrm{nm}$ in ZnO. It proves that we expect to have exceeded the Mott transition by far. In contrast to the simple comparison of only carrier densities to the Mott density, the estimate here accounts also in a simple manner for the high effective carrier temperatures.

The related Thomas-Fermi wave-vectors[3]

$$q_{\mathrm{TF}}^{\mathrm{e}} = \frac{1}{\lambda_{\mathrm{D}}^{\mathrm{e}}} \approx 1.7 \cdot 10^{9}\mathrm{m}^{-1} \,, \qquad\qquad q_{\mathrm{TF}}^{\mathrm{h}} = \frac{1}{\lambda_{\mathrm{D}}^{\mathrm{h}}} \approx 2.7 \cdot 10^{9}\mathrm{m}^{-1}$$

can be compared to the wave vectors corresponding to electrons and holes at the given effective temperatures assuming parabolic bands (cf. Ehrenreich [S13]):

$$q_{\mathrm{e}} = \sqrt{\frac{2m_{\mathrm{e}}k_{\mathrm{B}}T_{\mathrm{e}}}{\hbar^2}} \approx 1.9 \cdot 10^{9}\mathrm{m}^{-1} \,, \qquad\qquad q_{\mathrm{h}} = \sqrt{\frac{2m_{\mathrm{h}}k_{\mathrm{B}}T_{\mathrm{h}}}{\hbar^2}} \approx 1.9 \cdot 10^{9}\mathrm{m}^{-1} \,.$$

Non-parabolicity of the bands will increase these values significantly. As the estimated quasi Fermi-energies are within the band gap, these wave-vector values here can be understood as an equivalent to Fermi wave-vectors in the parabolic approximation. According to Ehrenreich, screening of the electron-phonon interaction would vanish if $q_{\mathrm{e/h}} \gg q_{\mathrm{TF}}^{\mathrm{e/h}}$. This is not the case within the parabolic approximation. Hence, it can be assumed that both, the electron-phonon coupling (Fröhlich coupling constant) is reduced and the LO phonon energy is shifted due to the free charge-carriers (red shift as lower phonon-plasmon branch for large plasma frequency). However, $q_{\mathrm{e/h}} \gg q_{\mathrm{TF}}^{\mathrm{e/h}}$ may become true when the nonparabolicity of the bands is taken into account. This could explain a suppression of the screening.

---

[3]The approximation of Eq. 4 in the article, as used for the zero-temperature excitonic correction, is not valid any more because $k_{\mathrm{B}}T_{\mathrm{e/h}} \ll |\tilde{E}_{\mathrm{F}}^{\mathrm{e/h}}|$ is not fulfilled. It holds $k_{\mathrm{B}}T_{\mathrm{e}} \approx 0.60\,\mathrm{eV}$ and $k_{\mathrm{B}}T_{\mathrm{h}} \approx 0.24\,\mathrm{eV}$.

# VIII. Comparison of the experimental dielectric function of excited ZnO with theoretical approaches

## VIII.I. First-principles simulations of excited electron-hole pairs at finite temperature

Complementary to Fig. 4 **d** of the main text, Fig. S11 shows the computed imaginary part of the DFs and, again, their differences between highly excited and unexcited state, along with experimentally obtained data. In the unexcited case (solid line and filled symbols in panel **b**), the discrepancy between model and experimental data arises mainly from the fact that the exciton absorption of the ZnO film is smeared out due to the surface pinning of the Fermi level. Hence, the computed DF overestimates the excitonic absorption of our film. In the excited case (dashed line and open symbols), the exciton absorption does not fully disappear although that would be expected from the computation. Furthermore, the theoretical model does not describe the EPC that we see in the experimental data. As discussed in the main text, contributions from phonon-assisted processes might also be the reason why the computed IVB is much smaller than the the experimentally observed one.

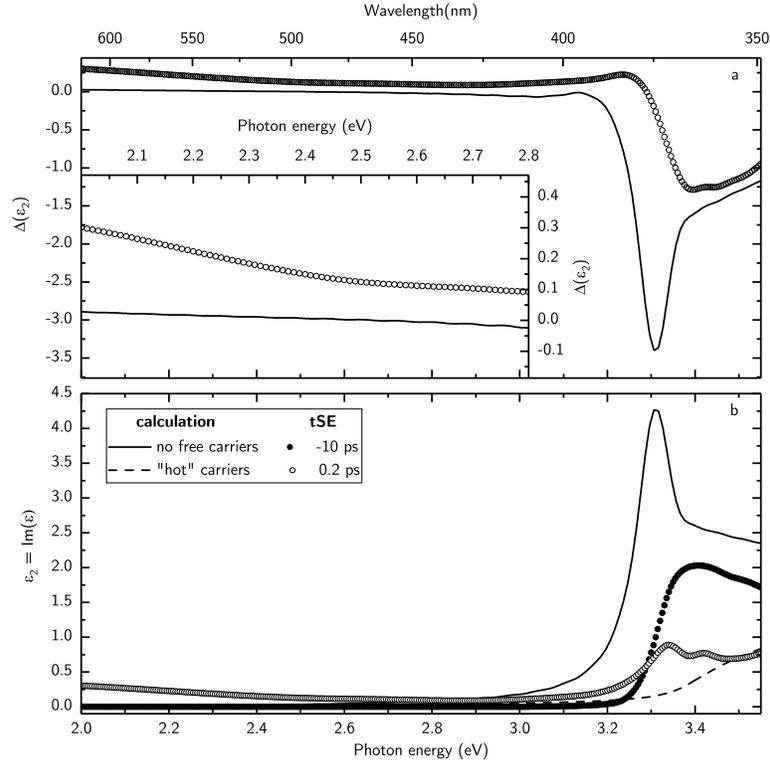

**Figure S11: Obtained versus computed $\varepsilon_2$ for high carrier excitation: a:** Experimental and computed difference of the imaginary part $\varepsilon_2$ of the DF for excited vs. non-excited ZnO. Symbols represent the difference from experimental data at -10 ps and 200 fs, lines the computed difference when assuming no or $10^{20}$ cm$^{-3}$ excited charge carriers with effective temperatures of 2800 K for holes and 7000 K for electrons. The inset shows a zoom into the IVB absorption range. **b:** Related $\varepsilon_2$ spectra.

## VIII.II. Existing models for highly excited ZnO and comparison to conventional transient spectroscopy

In comparison to ellipsometry, conventional reflectance and transmittance measurements lack any phase information of the electromagnetic waves interacting with the sample. This is usually compensated for by before-hand assumptions on the physical processes that, however, can lead to incorrect conclusions. Reflectance and transmittance spectra can be reconstructed from the knowledge of the DF. We generate reflectance spectra based on the DF obtained by time-resolved spectroscopic ellipsometry and compare them to theoretical values of Versteegh *et al.* [S12] which were refined by Wille *et al.* [S14]. The underlying DF of Wille *et al.* allows to explain gain and lasing mechanisms in ZnO micro- and nanowires [S15]. Both theoretical approaches are based on a solution of the Bethe-Salpeter equation [S16] for a simplified ZnO-like bulk system. The reflectance spectra are exemplary for various different pump-probe reflectance studies on ZnO [S17, 18, 19, 20]. Symbols in Fig. S12 show the DF as obtained in this work at selected pump-probe time delays; lines represent theoretical curves according to Wille *et al.* for various carrier densities. Both studies find a decrease in the real and the imaginary

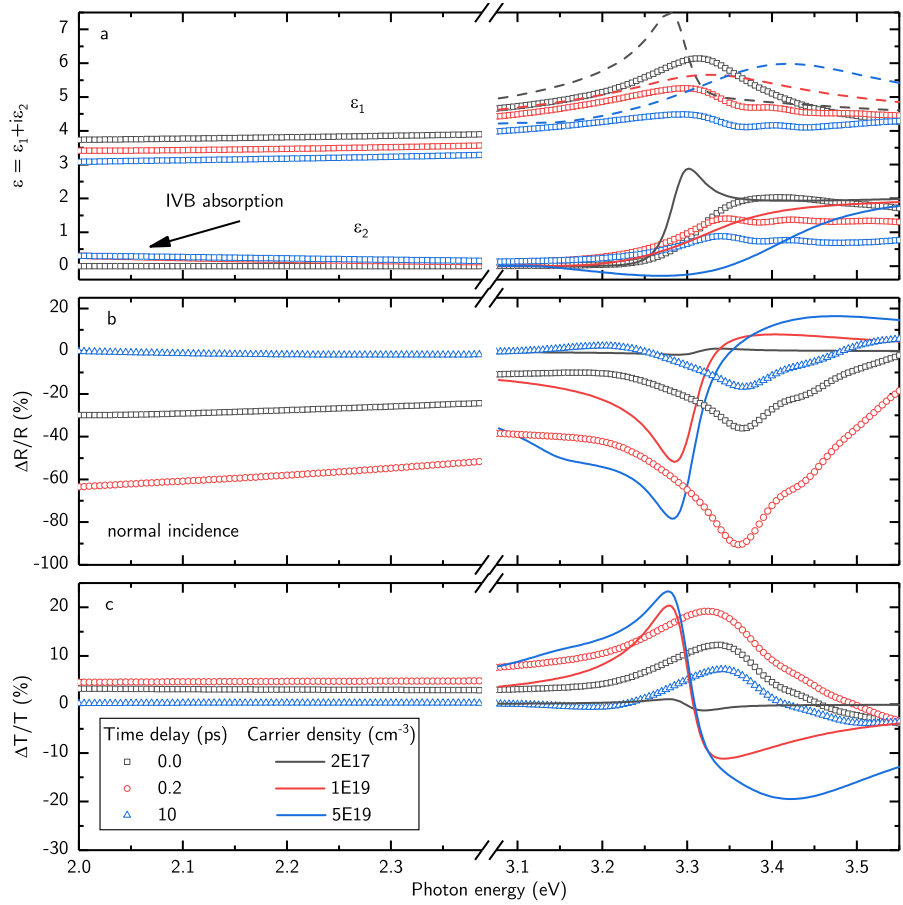

**Figure S12: Obtained DF vs. DF model and conventional spectroscopy:** **a**: DF of highly excited ZnO. Symbols represent the spectra obtained in this work at three different delays after photo-excitation. Lines (only for higher photon energies around the band edge) show the expected spectra according to the model of Wille *et al.* for three different charge-carrier densities. **b**: Computed transient reflectance and **c**: transmittance difference spectra at normal incidence for a 30 nm thin ZnO film on fused silica substrate according to the DF's in a. Note that although IVB absorption sets in, transmittance at lower energies increases upon pumping while reflectance decreases. This is caused by the lowered refractive index.

part of the DF with increasing carrier density. The model of Wille *et al.* is about 100 meV blueshifted and predicts $\varepsilon_2 < 0$ which can lead to optical gain and lasing. This is not observed in our experiment due to the reflection geometry. Optical gain can only occur due stimulated emission which produces photons of equal wave vector (magnitude and direction). So-called gain spectroscopy has only been reported in transmission geometry. Furthermore, it is seen that the theoretical curve of Wille *et al.* is not able to explain the features related to exciton-phonon complexes at 3.4 eV since electron-phonon interaction is neglected in the model. In the spectral range far below the band gap which is not covered by Wille *et al.*, we find increased absorption which is related to the IVB absorption.

The relative difference spectra of reflectance (panel b in Fig. S12) and transmittance (panel c) and are computed for a structure consisting of 30 nm *c*-plane oriented ZnO on a fused SiO$_2$ substrate which is equivalent to the sample studied in this work. Reflection from the substrate backside is ignored. Changes around the absorption edge of ZnO are on the same order of magnitude for both, using the DF from theoretical model (lines) and applying the DF obtained in this work. Surprisingly, in the spectral range of the IVB aborption the transmittance is **increased** although absorption appears. It is clear that the increased transmittance is related to decreased reflectance caused by the decrease in $\varepsilon_1$ and hence refractive index. This is in accordance with the Kramers-Kronig relations and is related to both, the occurring IVB absorption as well as the absorption bleaching at the absorption edge. We would like to emphasize here that interpretation of the conventional reflectance or transmittance changes can lead to erroneous conclusions about their physical origin because effects caused by changes in the real and imaginary part of the DF cannot be separated. Assuming a non-varying refractive index is insufficient and retrieval by exploiting the Kramers-Kronig relations is usually hampered by the limited spectral range.



\end{document}